\documentstyle[epsfig,cite]{mn2e}
\begin{document}

\title[Multiwavelength {\sl XMM-Newton} observations of the Laor et al. sample of PG quasars]
    {Multiwavelength {\sl XMM-Newton} observations of the Laor et al. sample of PG quasars}
\author[Brocksopp et al.]
    {C.~Brocksopp$^1$\thanks{email: cb4@mssl.ucl.ac.uk}, R.L.C.~Starling$^2$, P.~Schady$^1$, K.O.~Mason$^1$, E.~Romero-Colmenero$^3$,
\newauthor
 E.M.~Puchnarewicz$^1$\\
$^1$Mullard Space Science Laboratory, University College London, Holmbury St. Mary, Dorking, Surrey RH5 6NT\\
$^2$Astronomical Institute ``Anton Pannekoek'', University of Amsterdam, Kruislaan 403, 1098 SJ Amsterdam, The Netherlands\\
$^3$South African Astronomical Observatory, PO Box 9, Observatory 7935, Cape Town, South Africa\\
}
\date{Accepted ??. Received ??}
\pagerange{\pageref{firstpage}--\pageref{lastpage}}
\pubyear{??}
\maketitle

\begin{abstract}
We present {\sl XMM-Newton}/EPIC spectra for the Laor et al. sample of Palomar Green quasars. We find that a power-law provides a reasonable fit to the 2--5 keV region of the spectra. Excess soft X-ray emission below 2 keV is present for all objects, with the exception of those known to contain a warm absorber. A single power-law is, however, a poor fit to the 0.3--10.0 keV spectrum and instead we find that a simple model, consisting of a broken power-law (plus an iron line), provides a reasonable fit in most cases. The equivalent width of the emission line is constrained in just twelve objects but with low ($<2\sigma$) significance in most cases. For the sources whose spectra are well-fit by the broken power-law model, we find that various optical and X-ray line and continuum parameters are well-correlated; in particular, the power-law photon index is well-correlated with the FWHM of the {\sc H}$\beta$ line and the photon indices of the low and high energy components of the broken power-law are well-correlated with each other. These results suggest that the 0.3--10 keV X-ray emission shares a common (presumably non-thermal) origin, as opposed to suggestions that the soft excess is directly produced by thermal disc emission or via an additional spectral component. We present {\sl XMM-Newton} OM data which we combine with the X-ray spectra so as to produce broad-band spectral energy distributions, free from uncertainties due to long-term variability in non-simultaneous data. Fitting these optical--UV spectra with a Comptonized disc model indicates that the soft X-ray excess is independent of the accretion disc, confirming our interpretation of the tight correlation between the hard and soft X-ray spectra.
 
\end{abstract}


\section{Introduction}

\begin{table*}
\caption{Table listing the 23 PG quasars in the Laor et al. sample, their redshifts and Galactic absorption and details of the {\sl XMM-Newton} observations. Measurements of the Galactic absorption have been obtained from Laor et al. (1997 and references therein).}
\begin{tabular}{lccccccccc}
\hline
\hline
Quasar&Redshift&nH$_{\rm Gal.}$& Observation& {\sl XMM}& Observation & Exposure& Count Rate\\
& & ($\times10^{20}\,\,\mbox{cm}^{-2}$)&	Date&	Revolution&	ID&(PN/MOS; ks)&(PN, MOS1, MOS2; cts/s)\\
\hline	    
PG 0947+396& 0.2060	&1.92	&2001-11-03	&349&	0111290101&17.6/21.1 &  1.68     0.31   0.31      \\ 
PG 0953+414& 0.2341	&1.12	&2001-11-22	&358&	0111290201&10.6/13.2 &  3.64   	 0.63 	0.60      \\	
PG 1001+054& 0.1610     &2.88   &2003-05-04	&623&	0150610101&9.1/11.6  &  0.04   	 0.01 	0.01      \\ 
PG 1048+342& 0.1670	&1.74	&2002-05-13	&444&	0109080701&28.1/32.2 &  1.12   	 0.19 	0.19      \\	
PG 1114+445& 0.1440	&1.94	&2002-05-15	&445&	0109080801&37.8/42.3 &  0.66   	 0.18 	0.17      \\	
PG 1115+407& 0.1540	&1.74	&2002-05-17	&446&	0111290301&15.0/20.3 &  1.99   	 0.29 	0.30      \\	
PG 1116+215& 0.1765	&1.44	&2001-12-02	&363&	0111290401&5.5/8.4   &  4.79   	 0.74 	0.73      \\	
PG 1202+281& 0.1653	&1.72	&2002-05-30	&453&	0109080101&12.9/17.4 &  2.55   	 0.50 	0.50      \\	
PG 1216+069& 0.3313	&1.57	&2002-12-18	&554&	0111291101&14.0/16.5 &  1.03   	 0.19 	0.18      \\	
PG 1226+023& 0.1583	&1.68	&2000-06-13	&94&	0126700301&44.3/58.5 &  34.82  	 1.84 	1.36      \\	
PG 1309+355& 0.1840	&1.01	&2002-06-10	&458&	0109080201&25.2/29.0 &  0.43   	 0.08 	0.08      \\	
PG 1322+659& 0.1680	&2.88	&2002-05-11	&443&	0109080301&8.7/11.6  &  2.24   	 0.33 	0.33      \\	
PG 1352+183& 0.1520	&2.84	&2002-07-20	&478&	0109080401&12.5/14.9 &  2.19   	 0.34 	0.34      \\	
PG 1402+261& 0.1640	&1.42	&2002-01-27	&391&	0109081001&9.1/12.0  &  2.98   	 0.44 	0.44      \\	
PG 1411+442& 0.0896	&1.05	&2002-07-10	&473&	0103660101&27.5/41.1 &  0.11   	 0.02 	0.02      \\	
PG 1415+451& 0.1140	&0.96	&2002-12-08	&549&	0109080501&21.2/24.1 &  1.36   	 0.21 	0.22      \\	
PG 1425+267& 0.3660	&1.54	&2002-07-28	&482&	0111290601&45.9/--   &  0.63   	 0.45 	0.45      \\	
PG 1427+480& 0.2210	&1.69	&2002-05-31	&453&	0109080901&35.2/39.0 &  1.10   	 0.19 	0.20      \\	
PG 1440+356& 0.0790	&0.97	&2001-12-23	&373&	0107660201&24.8/30.4 &  6.26   	 0.78 	0.77      \\	
PG 1444+407& 0.2673	&1.09	&2002-08-11	&489&	0109080601&18.7/21.5 &  0.91   	 0.13 	0.13      \\	
PG 1512+370& 0.3707	&1.40	&2002-08-25	&496&	0111291001&17.6/20.4 &  1.45   	 0.28 	0.29      \\	
PG 1543+489& 0.4000     &1.68   &2003-02-08	&580&	0153220401&8.2/9.8   &  0.33   	 0.05 	0.05      \\	
PG 1626+554& 0.1330	&1.55	&2002-05-05	&440&	0109081101&5.5/8.9   &  2.92   	 0.44 	0.47      \\	
\hline 
\label{tab:quasars}
\end{tabular}
\end{table*}
 
The launch of {\sl XMM-Newton} has seen great advances in all areas of X-ray astronomy but its multiwavelength capabilities might be considered one of its greatest assets. This is of particular importance when it comes to studying quasars which are well-known for emitting at a broad range of frequencies. Different components of the quasar emit in different regions of the electromagnetic spectrum and so multiwavelength analysis is vital. 

The ``big blue bump'' (BBB) which is observed in the optical-ultraviolet range, peaking at $\sim 1000\, $\AA, is thought to be the signature of thermal emission from an accretion disc (e.g. Peterson 1997 and references therein). Above 2 keV the spectra can be well-modelled using a power-law, which is assumed to be due to Comptonization of lower energy photons. The ``soft excess'' above the power-law, often seen in X-ray spectra of quasars at energies below $\sim$ 2--3 keV, has been attributed to the high energy end of the BBB (although see also Gierli\'nski \& Done 2004); various models have been suggested for the origin of this soft X-ray excess such as direct thermal emission from the accretion disc or reprocessing of disc emission. (e.g. Peterson 1997 and references therein). Two additional components thought to be present in quasars, but not accessible to {\sl XMM-Newton}, are the infrared bump (assumed to be emission from dust) and the radio jet (e.g. Peterson 1997 and references therein).

The Laor et al. (1994, 1997) complete sample of 23 quasars was chosen from the Bright Quasar Survey (BQS), a subset of the Palomar Green (PG) survey compiled by Schmidt \& Green (1983). The PG survey consists of 114 AGN, 92 of which are quasars. With selection criteria $M_B<-23$ and $U-B<-0.44$, the BQS quasars are thought to be representative of most quasars (although see also Jester et al. 2005); since they were selected purely on their optical properties they were also thought to be unbiased in terms of their X-ray properties. In order to create a complete sample, the Laor et al. sample was further constrained by two additional selection criteria:

\begin{enumerate}
\item $z\le 0.400$, in order to prevent the rest frame 0.2 keV being redshifted beyond the range observable by {\sl ROSAT}/PSPC
\item $nH_{\rm Gal} <1.9\times 10^{20}\,\,\mbox{cm}^{-2}$, in order to minimize the effects of Galactic absorption
\end{enumerate} 

The full analysis of the {\sl ROSAT} observations was presented in Laor et al. (1997) and suggested that the 0.2--2 keV range of a quasar's rest frame spectrum could be fit by a single power-law, with no evidence for any soft excess emission and an upper limit (95\% confidence) of $\sim 5\times 10^{19}\,\,\mbox{cm}^{-2}$ on the amount of excess foreground absorption by cold gas. These data also suggested that the mean soft X-ray continuum agreed with an extrapolation of the mean UV continuum. The authors concluded that a thin bare accretion disc model was unable to reproduce the 0.2-2 keV spectral shape (Zheng et al. 1997; Laor et al. 1997).

We have used {\sl XMM-Newton} to update the {\sl ROSAT} observations, providing a much wider X-ray coverage (0.3--10 keV), higher energy resolution and a high S/N ratio, and also providing simultaneous optical and UV data from the Optical Monitor (OM). Our aims were to investigate whether or not claims to the presence of a soft excess, intrinsic absorption and/or iron lines were justified and physical when viewed in the context of the broad-band spectra. We have restricted our observations to the complete Laor et al. sample of 23 quasars but compare our X-ray spectral parameters with those obtained for alternative BQS quasars by Porquet et al. (2004; their sample consisted of 21 PG quasars, 15 of which were Laor objects) and Piconcelli et al (2005). In this paper we present the observations in Section 2, the X-ray spectroscopy in Section 3, line and continuum correlations in Section 4, the broad-band spectral energy distributions in Section 5 and discuss our results in Section 6.

\section{Observations}

{\sl XMM-Newton} observed the 23 Laor targets between 2001 November and 2003 May and obtained images with each of the three X-ray EPIC cameras (pn, MOS1, MOS2);  details of the sources and their observations are listed in Table~\ref{tab:quasars}. Each of the three X-ray cameras has a spectral range of 0.2--12 keV, spectral resolution of 20--50 and an angular resolution of $\sim 6 \arcsec$. The EPIC large window mode and thin filter were used (with the exception of PG~1411+442 and PG~1425+267 which were observed in full frame mode, and PG~1226+023 which was observed with the medium filter and small window mode in order to minimise pile-up). The data were reduced using the {\sc sas v5.4} {\sc epchain} and {\sc emchain} pipelines for the pn and MOS respectively. We re-reduced a selection of the sources with the {\sc sas v6.0} version of the software but found it made no significant difference to the subsequent fits. All datasets were inspected for pile-up by comparing the preliminary lightcurves with the published threshold count rates for each camera, as well as the {\sc epatplot} routine; PG~1226+023 ($\equiv$ 3C~273) was the only source affected by pile-up. Response matrices and auxiliary files were generated using the {\sc rmfgen} and {\sc arfgen} tasks respectively.

Longer wavelength images of each quasar were obtained, simultaneously with the X-ray data, using the Optical Monitor through the optical $U$, $B$ and $V$ filters and the UV $UVW1$, $UVM2$ and $UVW2$ filters (Mason et al. 2001). They were reduced using the {\sc sas v6.0} {\sc omichain} reduction pipeline, which uses an aperture of 12 pixels. We list the central wavelength of the filters and resultant photometry in Table~\ref{tab:photometry}. Finally the magnitudes were corrected for Galactic extinction using the extinction law of Cardelli et al. (1989) and flux calibrated with respect to Vega. Flux conversions were obtained using the forumla $M_{Vega}-M_{quasar}=-2.5\log(F_{Vega}/F_{quasar}$), where Vega has a brightness of 0.03 magnitudes in the $B$ and $V$ filters, 0.025 magnitudes in the $U$ and UV filters and fluxes of 3.2652743, 5.9712835, 3.1594819, 3.7391873, 4.4609810, 5.3884371 $\times 10^{-9}$ erg/cm$^2$/s/\AA~ for the $V$, $B$, $U$, $UVW1$, $UVM2$ and $UVW2$ filters respectively\footnote{http://xmm.vilspa.esa.es/sas/documentation/watchout/uvflux.shtml}.

\begin{table*}
\caption{Table of optical and UV photometry from the Optical Monitor observations. The second column under each filter heading gives the statistical error. Details of the OM and the filters are given in Mason et al. (2001). Flux conversions are stated in the text.}
\label{tab:photometry} 
\begin{tabular}{lcccccc}
\hline
\hline
&\multicolumn{6}{c}{Filter/magnitudes ($\lambda_{\rm eff}$)}\\
Quasar&$V$ (5483 \AA)&$B$ (4443 \AA)&$U$ (3735 \AA)&$UVW1$ (2910 \AA)&$UVM2$ (2310 \AA)&$UVW2$ (2120 \AA)\\
\hline
PG 0947+396&     --       &      --      &    15.31 0.01 &   15.13 0.01 & 14.92 0.01	& 14.77 0.01\\
PG 0953+414&     --       &      --      &        --     &   13.16 0.01 & 13.71 0.01	& 13.55 0.01\\
PG 1001+054&     --       &      --      &        --     &       --     &    --       	&     --    \\
PG 1048+342&  16.51 0.03  &   16.80 0.02 &    15.66 0.01 &   15.44 0.01 & 15.36 0.02	& 15.25 0.02\\
PG 1114+445&  15.78 0.01  &   16.17 0.01 &    15.08 0.01 &   15.00 0.01 & 15.15 0.01	& 15.08 0.02\\
PG 1115+407&     --       &   16.00 0.01 &    14.83 0.01 &   14.54 0.01 & 14.25 0.01	& 14.12 0.01\\
PG 1116+215&     --       &   14.32 0.01 &        --     &   12.85 0.01 &     --    	& 12.43 0.01\\
PG 1202+281&  16.27 0.03  &   16.52 0.02 &    15.32 0.01 &   15.10 0.02 & 14.98 0.02	& 14.80 0.03\\
PG 1216+069&  15.43 0.01  &   15.64 0.01 &    14.60 0.01 &   14.46 0.01 & 14.20 0.01	& 14.13 0.01\\
PG 1226+023&  12.64 0.01  &   12.94 0.01 &    11.77 0.01 &   11.49 0.01 & 11.33 0.01    & 11.28 0.01\\
PG 1309+355&  15.64 0.01  &   16.03 0.01 &    15.00 0.01 &   14.84 0.01 & 14.76 0.01	& 14.70 0.02\\ 
PG 1322+659&     --       &     --       &    14.82 0.01 &   14.58 0.01 & 14.36 0.01	& 14.27 0.02\\
PG 1352+183&  16.37 0.03  &   16.63 0.02 &    15.41 0.01 &   15.14 0.01 & 14.98 0.02	& 14.85 0.03\\ 
PG 1402+261&     --       &     --       &      --       &   14.04 0.01 & 13.74 0.01	& 13.61 0.01\\ 
PG 1411+442&     --       &     --       &    13.69 0.01 &       --     &   --          &   --      \\
PG 1415+451&  15.96 0.02  &   16.54 0.01 &    15.43 0.01 &   15.18 0.01 & 15.15 0.02	& 15.02 0.02\\
PG 1425+267&  17.16 0.04  &   17.17 0.02 &    16.05 0.01 &   16.10 0.01 & 15.90 0.02	& 15.79 0.03\\
PG 1427+480&  16.63 0.04  &      --      &    15.69 0.01 &   15.53 0.01 & 15.40 0.01	& 15.22 0.02\\ 
PG 1440+356&  14.63 0.01  &      --      &    13.92 0.01 &   13.66 0.01 & 13.54 0.01	& 13.43 0.01\\ 
PG 1444+407&  15.94 0.02  &   16.07 0.01 &    14.95 0.01 &   14.81 0.01 & 14.48 0.01	& 14.39 0.02\\
PG 1512+370&  16.65 0.03  &   16.75 0.01 &    15.67 0.01 &   15.58 0.01 & 15.06 0.02	& 14.85 0.02\\ 
PG 1543+489&       --     &       --   	 &    16.50 0.01 &       --     & 15.14 0.01    &     --    \\
PG 1626+554&  15.79 0.02  &       --     &         --    &   14.60 0.01 & 14.49 0.01	& 14.43 0.02\\ 
\hline
\end{tabular}
\end{table*}

\begin{table*}
\caption{Table of parameters resulting from a single power-law fit to the 2--5 keV energy range (rest frame; column 2 lists the equivalent observed frame energies), including only Galactic absorption. The penultimate column lists the luminosity of each quasar, assuming a standard WMAP cosmology. Clearly a power law spectrum is a reasonable assumption in this region of the spectrum for the majority of the sources. The final column shows the reduced $\chi^2$ for a power-law model fit to the full 0.3--10 keV range. Clearly this is a poor fit in all cases.}
\begin{tabular}{lccccccc}
\hline
\hline											   
Quasar&   Observed Frame &      Photon Index  		&$\chi^2$ (DOF)&$\chi^2_{\nu}$&Probability&   Luminosity&$\chi^2_{\nu}$\\
&Energy (keV)&&&&& ($\times 10^{44}$ erg$\,$s$^{-1}$)&(0.3--10 keV)\\
\hline
PG 0947+396&1.7 -- 4.1 & $2.10^{+0.08}_{-0.08}$ & 279.89 (286) & 0.98 & 0.59 & 1.18& 1.38\\
PG 0953+414&1.6 -- 4.1 & $2.19^{+0.07}_{-0.07}$ & 302.43 (290) & 1.04 & 0.30 & 2.89& 1.65\\
PG 1001+054&1.7 -- 4.3 & $0.14^{+0.46}_{-0.38}$ &   1.14  (8) & 0.14 & 1.00 & 0.025& 4.02\\
PG 1048+342&1.7 -- 4.3 & $1.96^{+0.09}_{-0.09}$ & 194.69 (213) & 0.91 & 0.81 & 0.55& 1.45\\
PG 1114+445&1.7 -- 4.4 & $1.36^{+0.06}_{-0.05}$ & 412.14 (443) & 0.93 & 0.85 & 0.57& 4.64\\
PG 1115+407&1.7 -- 4.3 & $2.34^{+0.10}_{-0.10}$ & 170.59 (192) & 0.89 & 0.86 & 0.50& 1.55\\
PG 1116+215&1.7 -- 4.2 & $2.22^{+0.09}_{-0.09}$ & 226.40 (217) & 1.04 & 0.32 & 1.79& 1.51\\
PG 1202+281&1.7 -- 4.3 & $1.76^{+0.07}_{-0.07}$ & 290.70 (342) & 0.85 & 0.98 & 1.36& 1.54\\
PG 1216+069&1.5 -- 3.8 & $2.02^{+0.12}_{-0.12}$ & 184.09 (142) & 1.30 & 0.01 & 2.14& 1.97\\
PG 1226+023&1.7 -- 4.3 & $1.64^{+0.01}_{-0.01}$ &1425.47 (864) & 1.65 & 1e-13 & 27.0& 4.99\\
PG 1309+355&1.7 -- 4.2 & $1.80^{+0.11}_{-0.11}$ & 121.05 (137) & 0.88 & 0.83 & 0.34& 1.74\\
PG 1322+659&1.7 -- 4.3 & $2.34^{+0.13}_{-0.13}$ & 120.17 (117) & 1.03 & 0.40 & 0.64& 1.48\\
PG 1352+183&1.7 -- 4.3 & $2.10^{+0.11}_{-0.11}$ & 142.59 (146) & 0.98 & 0.56 & 0.66& 1.39\\
PG 1402+261&1.7 -- 4.3 & $2.42^{+0.10}_{-0.10}$ & 185.92 (189) & 0.98 & 0.55 & 0.82& 1.63\\
PG 1411+442&1.8 -- 4.6 & $0.34^{+0.19}_{-0.31}$ &  46.84  (40) & 1.17 & 0.21 & 0.02& 7.33\\
PG 1415+451&1.8 -- 4.5 & $2.14^{+0.09}_{-0.09}$ & 174.80 (195) & 0.90 & 0.85 & 0.21& 1.68\\
PG 1425+267&1.5 -- 3.7 & $1.49^{+0.07}_{-0.07}$ & 436.24 (379) & 1.15 & 0.02 & 3.05& 1.47\\
PG 1427+480&1.6 -- 4.1 & $2.04^{+0.07}_{-0.07}$ & 296.92 (327) & 0.91 & 0.88 & 0.89& 1.45\\
PG 1440+356&1.9 -- 4.6 & $2.41^{+0.06}_{-0.07}$ & 315.64 (363) & 0.87 & 0.96 & 0.26& 2.24\\
PG 1444+407&1.6 -- 3.9 & $2.37^{+0.14}_{-0.13}$ & 141.24 (116) & 1.22 & 0.06 & 0.76& 1.45\\
PG 1512+370&1.5 -- 3.6 & $1.95^{+0.08}_{-0.08}$ & 247.66 (240) & 1.03 & 0.35 & 4.69& 1.33\\
PG 1543+489&1.4 -- 3.6 & $2.54^{+0.35}_{-0.26}$ &  29.16  (33) & 0.88 & 0.66 & 0.71& 1.10\\
PG 1626+554&1.8 -- 4.4 & $2.01^{+0.10}_{-0.10}$ &  30.45  (39) & 0.78 & 0.83 & 0.85& 1.68\\
\hline	   
\label{tab:2-5kev}
\end{tabular}
\end{table*}

\section{X-ray Spectroscopy}

\begin{figure*}
\begin{center}
\leavevmode
\hspace*{-1cm}\epsfig{file=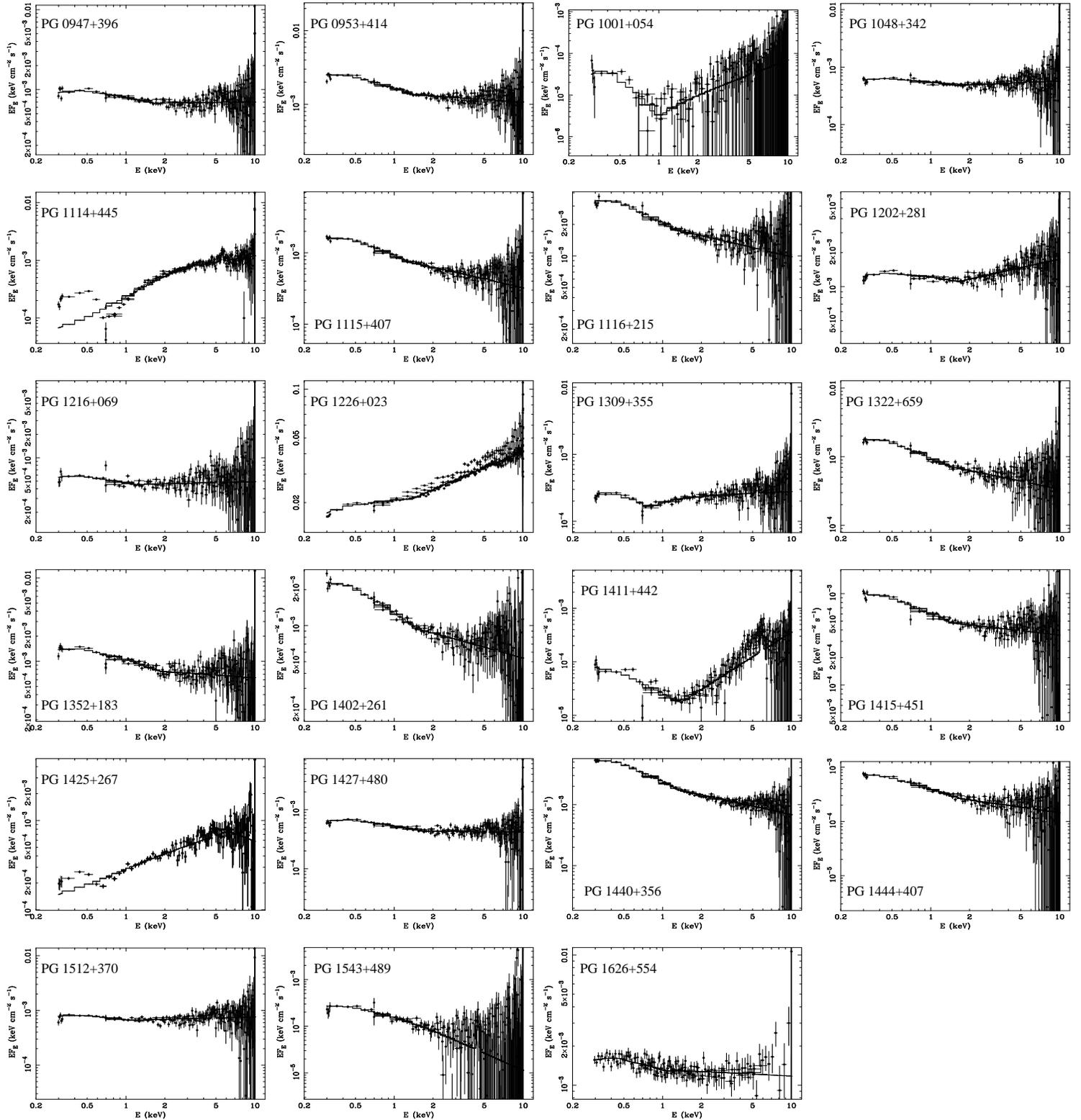}
\caption{A plot for each quasar showing the X-ray spectra (pn and MOS) in $\nu F_{\nu}$-space and the best fit to a model comprising an absorbed (allowing for both Galactic + intrinsic absorption) broken power-law plus iron line. This simple model provides a good fit to the spectra for the majority of the sources. We note that Ashton et al. (2004) have also included a warm absorber component in order to fit PG~1114+445 and PG~1309+355 successfully and suspect that similar detailed modelling would be required in order to fit the spectra of PG~1425+267 and PG~1411+442 (Ashton, Brocksopp et al. in prep.).}
\label{fig:febknpl}
\end{center}
\end{figure*}

\begin{table*}
\caption{Model-fit to the 0.3--10 keV (observed frame) data using an absorbed broken power-law with the addition of a single iron line of width 0.01 keV in the 5.5--8.0 keV (rest frame) range. The observed luminosity in this energy range has also been listed. This model provides a reasonably good fit for the majority of sources with the most notable exceptions being PG~1114+445 (see Ashton et al. 2004) and PG~1411+442 (which is unconstrained; see Ashton, Brocksopp et al. in prep).}
\begin{tabular}{lccccccccc}
\hline
\hline
Quasar& PhoInd. 1& PhoInd. 2& Break E&Absorption&Cont. Norm.&Luminosity&$\chi^2$ (DOF)&$\chi^2_{\nu}$& Probability\\
& & &  (keV) &($\times10^{20}\,\,\mbox{cm}^{-2}$)&($\times 10^{-3}$)&($\times 10^{44}\,\,\mbox{ergs\,\,s}^{-1}$)&&&\\
\hline

PG 0947+396& $2.43^{+0.07}_{-0.03}$ & $1.87^{+0.13}_{-0.09} $ & $2.02^{+0.30}_{-0.44}$ & $<0.63$  		& 0.84 	& 5.4  &  786.29 (790)  &1.00&0.53  \\
PG 0953+414& $2.65^{+0.12}_{-0.03}$ & $2.07^{+0.06}_{-0.08} $ & $1.56^{+0.22}_{-0.19}$ & $<1.54$  		& 1.69 	& 15.7 &  826.15 (777)  &1.06&0.11  \\
PG 1001+054& $4.36^{+1.72}_{-0.54}$ & $0.74^{+0.40}_{-0.39} $ & $1.03^{+0.21}_{-0.16}$ & $<3.35$		& 0.01  & 0.1  &  27.79 (26)	&1.07&0.37  \\
PG 1048+342& $2.39^{+0.11}_{-0.04}$ & $1.84^{+0.06}_{-0.09} $ & $1.66^{+0.30}_{-0.14}$ & $<1.56$  		& 0.56 	& 2.0  &  714.16 (675)  &1.06&0.14  \\
PG 1114+445& --                     & --                      &  --                    &    --                  &  --   &  --  &  --            &  --&--    \\
PG 1115+407& $2.99^{+0.17}_{-0.11}$ & $2.36^{+0.06}_{-0.17} $ & $1.35^{+0.22}_{-0.13}$ & $<2.75$  		& 0.96 	& 3.3  &  608.14 (649)  &0.94&0.87  \\
PG 1116+215& $2.76^{+0.19}_{-0.04}$ & $2.19^{+0.08}_{-0.08} $ & $1.46^{+0.22}_{-0.22}$ & $<1.96$  		& 2.13 	& 10.3 &  702.85 (677)  &1.04&0.24  \\
PG 1202+281& $2.27^{+0.04}_{-0.03}$ & $1.71^{+0.05}_{-0.05} $ & $1.70^{+0.14}_{-0.17}$ & $<0.50$  		& 1.24 	& 5.5  &  889.89 (902)  &0.98&0.61  \\
PG 1216+069& $3.09^{+0.32}_{-0.20}$ & $1.97^{+0.08}_{-0.08} $ & $0.96^{+0.17}_{-0.06}$ & $4.25^{+7.55}_{-2.59}$ & 0.39  & 2.3  &  655.40 (480)  &1.37&2e-7  \\
PG 1226+023& $2.04^{+0.01}_{-0.01}$ & $1.68^{+0.01}_{-0.01} $ & $1.34^{+0.03}_{-0.04}$ & $<0.12$                &21.52  &94.4  & 4353.74 (2785) &1.56&$<$1e-13  \\
PG 1309+355& $3.69^{+0.54}_{-0.51}$ & $1.83^{+0.07}_{-0.06} $ & $0.78^{+0.02}_{-0.04}$ & $6.77^{+4.70}_{-2.52}$ & 0.13  & 1.3  &  456.39 (473)  &0.96&0.70  \\
PG 1322+659& $3.03^{+0.19}_{-0.09}$ & $2.24^{+0.09}_{-0.12} $ & $1.49^{+0.15}_{-0.16}$ & $<2.76$  		& 1.00 	& 4.3  &  476.39 (495)  &0.96&0.72  \\
PG 1352+183& $2.61^{+0.05}_{-0.04}$ & $1.99^{+0.11}_{-0.07} $ & $1.91^{+0.24}_{-0.25}$ & $<0.55$  		& 1.07 	& 3.5  &  587.43 (561)  &1.05&0.21  \\
PG 1402+261& $2.87^{+0.11}_{-0.04}$ & $2.23^{+0.09}_{-0.10} $ & $1.55^{+0.23}_{-0.14}$ & $<1.31$  		& 1.31 	& 5.4  &  727.0 (642)  &1.13 &0.01  \\
PG 1411+442& $3.37^{+0.47}_{-0.15}$ & $0.35^{+0.11}_{-0.10} $ & $1.34^{+0.09}_{-0.12}$ & $<0.04$                & 0.02  & 0.1  &  308.3 (157)  &1.96 &6e-12  \\
PG 1415+451& $2.69^{+0.08}_{-0.03}$ & $2.03^{+0.09}_{-0.07} $ & $1.66^{+0.16}_{-0.21}$ & $<0.79$  		& 0.62 	& 1.2  &  698.40 (671)  &1.04&0.23  \\
PG 1425+267& $3.09^{+0.93}_{-0.39}$ & $1.49^{+0.04}_{-0.10} $ & $0.67^{+0.05}_{-0.04}$ & $4.69^{+8.33}_{-3.31}$ & 0.16  & 9.9  &  438.37 (409)  &1.07&0.56  \\
PG 1427+480& $2.48^{+0.04}_{-0.03}$ & $1.96^{+0.07}_{-0.07} $ & $1.81^{+0.20}_{-0.26}$ & $<1.27$  		& 0.55 	& 4.2  &  821.74 (843)  &0.97&0.69  \\
PG 1440+356& $3.18^{+0.12}_{-0.09}$ & $2.43^{+0.07}_{-0.06} $ & $1.43^{+0.13}_{-0.13}$ & $1.15^{+1.01}_{-0.79}$ & 2.50  & 2.2  &  1007.30 (921) &1.09&0.02  \\
PG 1444+407& $2.84^{+0.11}_{-0.04}$ & $2.07^{+0.21}_{-0.20} $ & $2.00^{+0.37}_{-0.46}$ & $<1.34$  		& 0.40 	& 5.2  &  480.77 (460)  &1.05&0.24  \\
PG 1512+370& $2.33^{+0.18}_{-0.04}$ & $1.85^{+0.09}_{-0.06} $ & $1.41^{+0.21}_{-0.34}$ & $<1.30$  		& 0.71 	& 19.1 &  679.47 (663) &1.02 &0.32  \\
PG 1543+489& $3.87^{+1.79}_{-0.67}$ & $2.85^{+0.34}_{-0.80} $ &	$0.90^{+2.25}_{-0.20}$ & $<32$		        & 0.15	& 4.6  &  164.93 (158) &1.04 &0.34  \\
PG 1626+554& $2.44^{+0.24}_{-0.10}$ & $2.05^{+0.06}_{-0.13} $ & $1.05^{+0.57}_{-0.16}$ & $<2.24$ 		& 1.36 	& 3.6  &  180.47 (166)  &1.09&0.21  \\
\hline
\end{tabular}
\label{tab:febknpl}
\end{table*}

\begin{table*}
\caption{Equivalent widths of a single possible iron line, of width 0.01 or 1.0 keV and rest frame energies 6.4, 6.5 or 6.7 keV, of an absorbed broken power-law model. Upper limits on the equivalent width are to within 99\%. We note that Page et al. (2004a) detected a weak, broad line in PG~1226+023 on co-adding a number of different observations.$^{\dagger}$The equivalent width for PG~1114+445 is obtained from Ashton et al. (2004) who include a warm absorber in the model.} 
\begin{tabular}{lcccccc}
\hline
\hline
Quasar& EW (0.01 keV)&EW (0.01 keV) & EW (0.01 keV)& EW (1.0 keV)&EW (1.0 keV) & EW (1.0 keV)\\
      & 6.4 keV &6.5 keV & 6.7 keV& 6.4 keV &6.5 keV & 6.7 keV\\
\hline

PG~0947+396&	$93^{+68}_{-65}$        &$89^{+69}_{-65}$	&$<108$                &      $203^{+202}_{-192}$         &     $205^{+ 206}_{- 195}$	    &$205^{+ 215}_{-  204}$	        \\
PG~0953+414&	$<62$	 		&$<70$			&$<121$  	       &	     $<196$	     	  &    	$<202$	    		    &$206^{+ 211}_{-  203}$	        \\
PG~1001+054&	$<2$	 		&$<2$			&$<2$                  &	     $<10^6$    	  &    	$<10^6$   		    &$<10^6$     \\
PG~1048+342&	$<124$ 			&$<103$			&$<165$  	       &	     $236^{+246}_{-233}$  &    	$<252$   		    &$<261$	        \\
PG~1114+445$^{\dagger}$& 	--	&$230^{+90}_{-70}$                  	&--	   	       &	     --	     	 	  &    	--	    		    &--	        \\
PG~1115+407&	$<155$ 			&$<166$			&$125^{+126}_{-121}$   &	     $462^{+324}_{-318}$  &    	$488^{+ 336}_{- 329}$	    &$534^{+ 358}_{-  352}$	        \\
PG~1116+215&	$<99$ 			&$<88$			&$101^{+102}_{-97}$    &	     $300^{+270}_{-259}$  &    	$316^{+ 278}_{- 267}$	    &$340^{+ 293}_{-  281}$	        \\
PG~1202+281&	$<84$ 			&$<61$			&$<43$		       &	     $<16$	     	  &     $<172$	    		    &$<180$	        \\
PG~1216+069&	$<168$ 			&$<152$			&$<172$		       &	     $298^{+311}_{-265}$  &    	$310^{+ 318}_{- 271}$	    &$332^{+ 333}_{-  284}$	    \\   
PG~1226+023&	$9^{+6}_{-7}$		&$<12$			&$9^{+7}_{-7}$	       &	     $53^{+24}_{-24}$	  &    	$51^{+ 25}_{- 25}$	    &$44^{+ 26}_{-  26}$	        \\
PG~1309+355&	$137^{+155}_{-116}$	&$<223$			&$<224$ 	       &	     $<382$	     	  &    	$<385$	    		    &$<394$	        \\    
PG~1322+659&	$<280$ 			&$178^{+178}_{-159}$	&$<273$  	       &	     $<431$	     	  &    	$<447$    		    &$<473$	        \\  
PG~1352+183&	$<187$ 			&$<186$			&$<194$ 	       &	     $<329$	     	  &    	$<337$   		    &$<352$	        \\
PG~1402+261&	$<186$			&$<169$			&$<162$ 	       &	     $446^{+312}_{-302}$  &    	$481^{+ 321}_{- 313}$	    &$544^{+ 342}_{-  334}$	        \\
PG~1411+442&	$<706$           	&$<683$	                &$<428$ 	       &	     $<706$		  &    	$<683$		    	    &$<428$		        \\
PG~1415+451&	$<183$ 			&$<170$			&$117^{+115}_{-104}$   &	     $<300$     	  &    	$<307$	    		    &$<319$	        \\
PG~1425+267&	$<113$	 		&$<146$			&$<114$		       &	     $<213$	     	  &    	$<214$	    		    &$<209$	        \\
PG~1427+480&	$<110$ 			&$<128$			&$<108$		       &	     $<183$     	  &    	$<188$	    		    &$<195$	        \\    
PG~1440+356&	$<136$	 		&$<152$			&$<131$		       &	     $424^{+246}_{-240}$  &    	$449^{+ 254}_{- 250}$	    &$494^{+ 273}_{- 268}$	    \\  
PG~1444+407&	$<150$ 			&$<197$			&$<217$		       &	     $<364$	     	  &    	$<370$	    		    &$<386$	        \\
PG~1512+370&	$74^{+69}_{-59}$	&$81^{+71}_{-62}$	&$<131$		       &	     $210^{+205}_{-176}$  &    	$219^{+ 208 }_{-180}$	    &$231^{+ 217}_{-  186}$	    \\
PG~1543+489&	$<169$	 		&$<174$			&$<204$	     	       &	     $<10^6$   	 	  &    	$<10^6$		            &$<10^6$    \\
PG~1626+554&	$<344$	 		&$<401$			&$<376$		       &	     $<486$               &    	$<512 $        		    &$<572$             \\
\hline
\end{tabular}
\label{tab:line}
\end{table*}

The task {\sc xmmselect} was used to extract an X-ray spectrum from a circular region centred on the source. Further circular regions were positioned on source-free areas of sky to provide a background sample. We corrected for the pile-up in PG~1226+023 by running the {\sc sas v6.0} chains and extracting the spectrum from an annulus centered on the centre of the source. Photons with pattern 0--4 and 0--12, for the pn and MOS respectively, and with quality flag 0 were included. {\sc grppha} was used to bin the data using a minimum of 20 counts in each bin and 150 counts per bin in the case of PG~1626+554. The extracted spectra were analysed using the {\sc xspec} spectral fitting package (Arnaud et al. 1996).

Initial fits were made using a simple absorbed power-law, taking into account both Galactic absorption (obtained from Laor et al. 1997 and references therein) and possible redshifted absorption intrinsic to the quasar. We restricted the energy range to 0.3--10 keV in the observed frame for which the instrument response is generally well-known. It was clear however that there was a significant discrepancy between the pn and MOS spectra with as much as a 10\% higher value of the power-law photon index for the pn relative to that of the MOS. Kirsch (2003) outlines two possible explanations for this problem; firstly a gradual change in the MOS response from revolution 200, and particularly 450, onwards (as discussed by Porquet et al. 2004). The second factor is a $\sim10$\% higher flux for the pn in the 0.3--1.0 keV range compared with the MOS, which is possibly due to uncertainties in the vignetting and CCD-quantum-efficiency. Since our sources are all relatively soft we suggest that both factors are contributing to the discrepancy. While the first factor is an uncertainty with the MOS calibration, we note that the second factor is not necessarily restricted to the MOS. We therefore do not discard all the MOS data (contrary to Porquet et al. 2004) but discard only the MOS data below 0.7 keV. We find that the pn and MOS are in good agreement (typically to within 90\% confidence contours) above this energy. We re-reduced four sources chosen at random using the {\sc sas v6.0} software and found that there was no significant improvement in pn/MOS agreement.

We then performed power-law fits to the spectra, fitting pn and MOS simultaneously and restricting the energy range to 2--5 keV in the rest frame (in contrast to Porquet et al. 2004). We included only the effect of Galactic absorption and note that this region of the spectrum is useful for preliminary fits, since it is typically thought to be free of model components due to soft excess, absorption, reflection or emission lines. A power-law was indeed a good fit to the spectra in this range and the resultant parameters are listed in Table~\ref{tab:2-5kev}. The values of $\Gamma$ fall in the range 1.4--2.6, with the exception of PG~1001+054 and PG~1411+442. Extrapolating the 2--5 keV fit to the full {\sl XMM} range confirms the result of Porquet et al. (2004) that a soft excess is present in all cases, except for those complicated by the presence of a warm absorber (Ashton et al. 2004; Ashton, Brocksopp et al. in prep.).

Indeed, over the full {\sl XMM-Newton} energy range (0.3--10 keV) a simple absorbed power law is not a good fit to the data (see $\chi^2_{\nu}$ in final column of Table~\ref{tab:2-5kev}). Many of the sources show excess emission in the low and/or high energy regions of the spectrum. Again, this confirms the results of Porquet et al. (2004) and Piconcelli et al. (2004) and so we do not repeat the work here. Instead we try fitting a broken power-law, again including Galactic absorption and possible intrinsic, redshifted absorption (Fig.~\ref{fig:febknpl}). This gives a more acceptable fit ($0.94\le \chi^2_{\nu} \le 1.13$), with the exception of PG~1114+445, PG~1216+069, PG~1226+023 and PG~1411+442 as discussed below (Table~\ref{tab:febknpl}). Intrinsic broad-band absorption is not required in the majority of our sources, although Porquet et al. (2004) included narrow absorption edges for some objects. Values of the photon index for the low energy power-law fall in the 2.0--4.0 range; for the high energy power-law the photon indices lie in the range 1.5--2.9 (again with the exceptions of PG~1001+054 and PG~1411+442). The break energy was in the range 0.7--2.1 keV (or 0.9--2.5 keV in the rest frame).

We also include a single iron line of width 0.01 or 1.0 keV in our model, placing it at each of the three energies 6.4, 6.5 and 6.7 keV  (Table~\ref{tab:line}). However, we find that the equivalent widths of any potential narrow or broad lines can be constrained in just twelve sources; with most of these detections having a significance of $<2\sigma$, it is debateable as to whether these lines can be considered real. 99\% confidence upper limits to the equivalent width are also listed in Table~\ref{tab:line}. The central energies of the detected lines are typically consistent with 6.4 or 6.5 keV, corresponding to neutral and weakly ionized iron emission. A more detailed search for iron emission in PG~1226+023 has been published by Page et al. (2004a) and a weak broad line found by co-adding spectra. We note that Jim\'enez-Bail\'on et al. (2005) study a sample of 38 PG quasars and detect iron lines in 20 of them, 13 of which are included in the Laor et al. sample. This does not necessarily contradict our results as these authors use a variety of different models, some of which we show in Sections 4--6 may be inappropriate. Further discussion of the iron lines in these sources, with reference to reflection models, can also be found in Porquet et al. (2004).

Alternative models, such as black-body, bremsstrahlung, Comptonization and more complicated absorption components, are often used to improve fits. Porquet et al. (2004) and Piconcelli et al. (2004) have performed detailed fitting of these models to sub-sets of the Laor et al. sample and so we do not repeat their work. These authors found that the addition of black-body or bremsstrahlung model components improved fits but that no single model was able to fit all the objects. The values of the resultant parameters were also regarded as phenomenological rather than physical, in particular requiring temperatures much higher than those thought to be reached in an accretion disc (Piconcelli et al. 2004). Comptonization models tended to be more successful and provided a good fit to the soft excess of the PG quasars (Porquet et al. 2004; Piconcelli et al. 2004; Gierli\'nski \& Done 2004). While the Comptonization of disc photons is indeed a potential physical model, there are still problems in that the resultant temperatures lie in an unrealistically narrow range (0.1--0.2 keV), which does not reflect the wider range of probable disc temperatures, and the soft X-ray emitting region would have to be extremely compact ($R\sim10^{12}$ cm).

The few sources which are not well-described by the simple broken power-law model merit further discussion. Ashton et al. (2004) perform detailed modelling of PG~1114+445 and PG~1309+355 and require the addition of a warm absorber component. PG~1425+267 is qualitatively similar to PG~1309+355 and may also require an additional absorption component (as suggested previously by Reeves \& Turner 2000). PG~1001+054 and PG~1411+442 are significantly different from the other sources; PG~1411+442 in particular appears to require an additional high energy component, perhaps reflection which would be consistent with the apparent strong emission line present in this source. Broken power-law fits to the spectra of PG~1216+069 and PG~1226+023 are also unacceptable, although in these cases it is not so clear why; the PG~1226+023 spectra are of much higher S/N than the other sources and this seems to be the reason why the fit is not as successful (suggesting, perhaps, that higher S/N observations of the other sources would also require more sophisticated models). Detailed modelling of individual sources is beyond the scope of this paper but will be addressed in follow-up work.

\begin{figure*}
\begin{center}
\leavevmode
\hspace*{-1cm}\epsfig{file=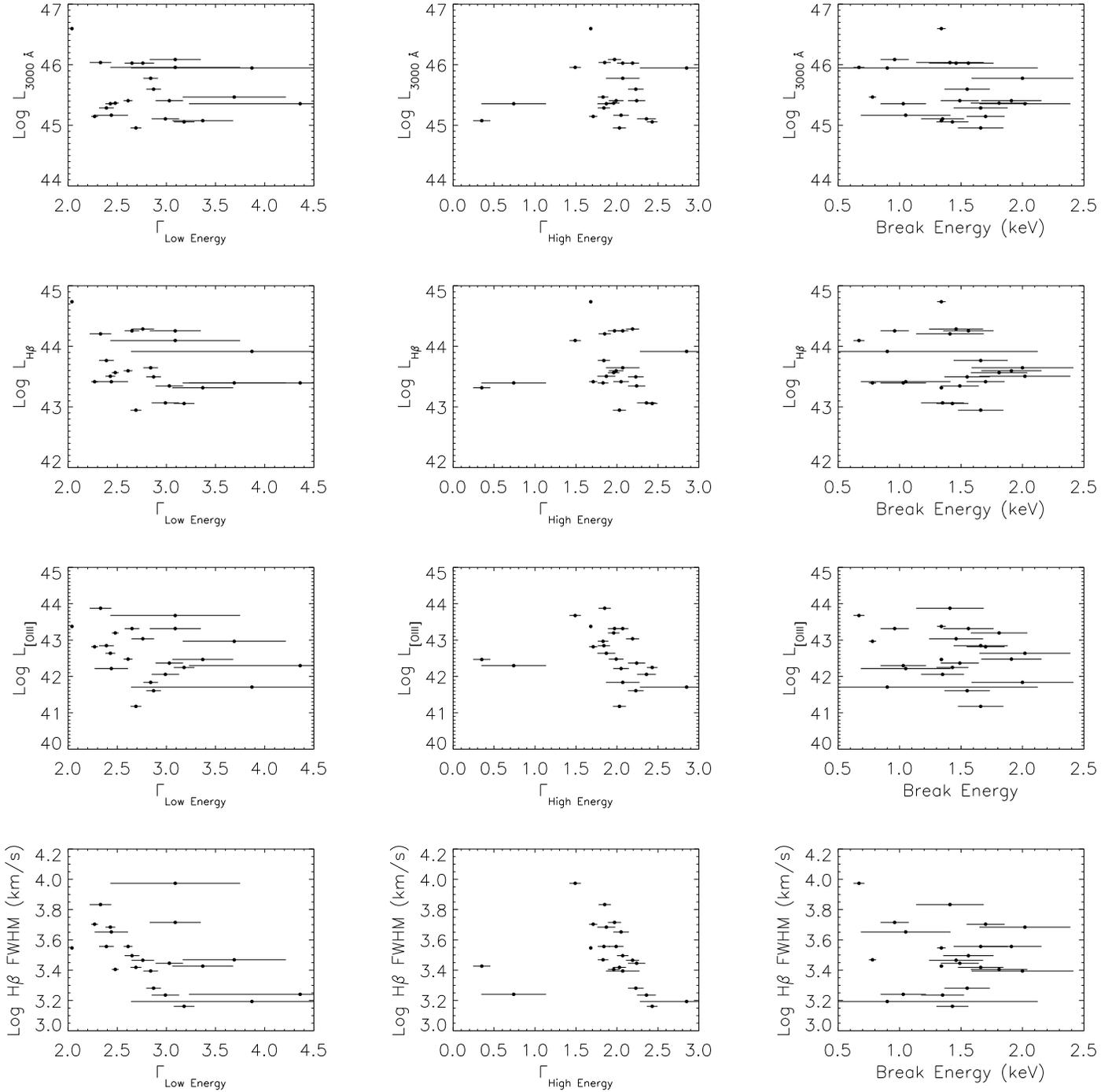}
\caption{Scatter plots showing various optical parameters (obtained from Laor et al. (1997)) plotted against the broken power-law parameters obtained in Section 3 (omitting PG~1114+445). The break energies show no indication of correlated behaviours; however, with the exception of PG~1001+054 and PG~1411+442, the [{\sc OIII}] luminosity and, particularly, the {\sc H}$\beta$ FWHM are correlated with the photon index, in the sense that as the spectrum hardens the [{\sc OIII}] luminosity and {\sc H}$\beta$ FWHM increase. Spearman rank order correlation coefficients for each plot are listed in Table~\ref{tab:correlation}.}
\label{fig:optical-pi}
\end{center}
\end{figure*}

\begin{figure*}
\begin{center}
\leavevmode
\hspace*{-1cm}\epsfig{file=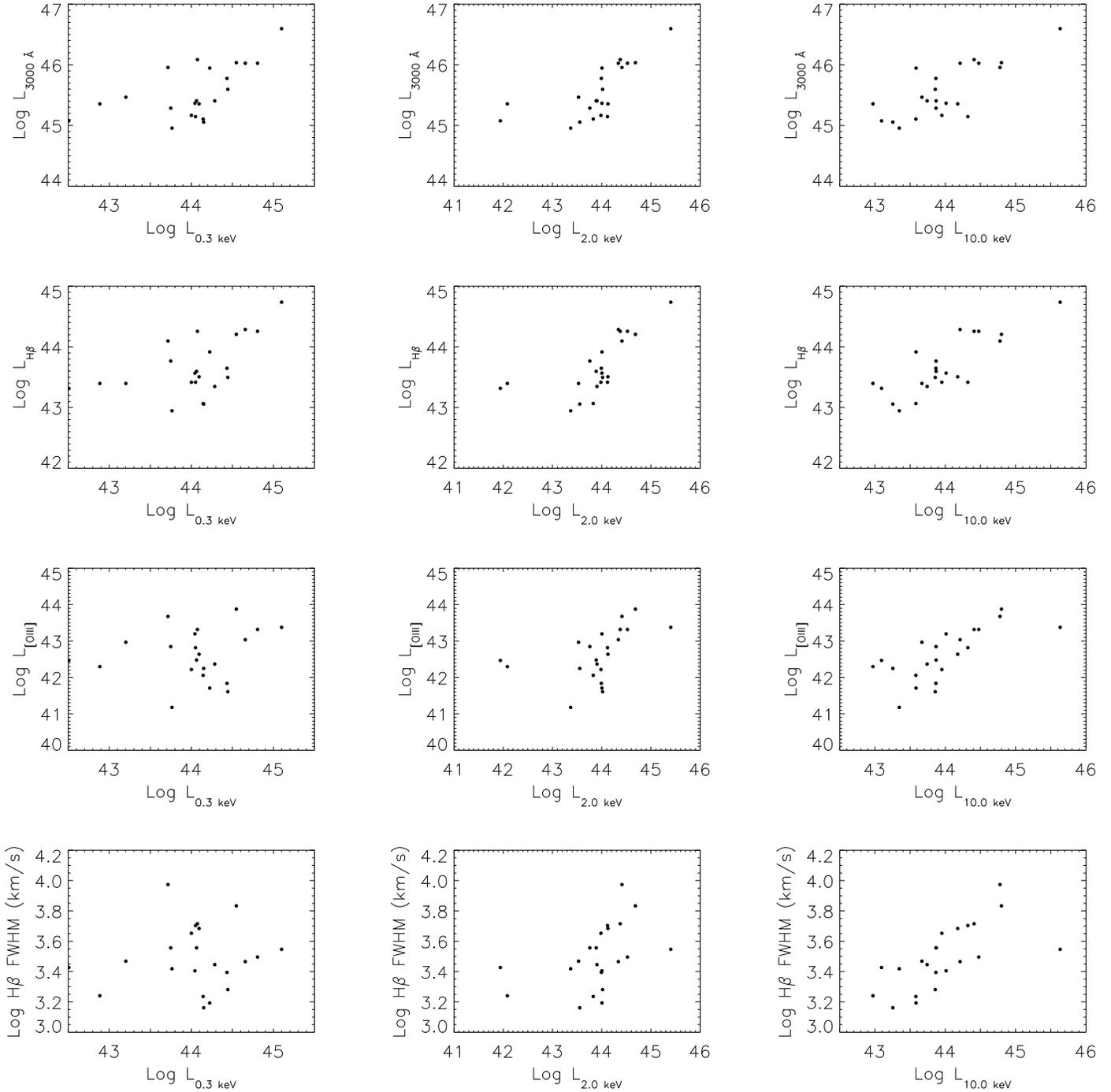}
\caption{Scatter plots showing various optical parameters (obtained from Laor et al. (1997)) plotted against X-ray luminosities at 0.3, 2.0 and 10.0 keV (omitting PG~1114+445). It is clear that as the 2.0 and 10.0 keV luminosities increase, so too do the values of all four optical parameters; any correlation at 0.3 keV is less convincing, perhaps on account of the EPIC response. Pearson correlation coefficients for each plot are listed in Table~\ref{tab:correlation}.}
\label{fig:optical-xray}
\end{center}
\end{figure*}

\section{X-ray/optical continuum and line luminosity correlations}

We have plotted various combinations of spectral parameters and line and continuum\footnote{Here we use only the luminosity derived using the pn data since the MOS spectra were unreliable at 0.3 keV} luminosities in Figures~\ref{fig:optical-pi} and \ref{fig:optical-xray}. The optical data were obtained from Laor et al. (1997) but we note that the lack of simultaneous observations of any varying sources is more likely to weaken the correlations than to strengthen them. We then calculated the associated Spearman rank order correlation coefficients ($\rho$); many of the plots show only scatter, but there are also a number of significant correlations ($0.6 \le \rho < 0.9$ to within $99\%$ confidence). We list these higher values of $\rho$ in Table~\ref{tab:correlation}. We note that PG~1114+445 has been omitted from the plots on account of its extremely poor fits to the broken power-law model. For each pair of parameters in the table, the first value of $\rho$ is for the plots from Figures~\ref{fig:optical-pi} and \ref{fig:optical-xray}; the second value is obtained through further omission of PG~1001+054, PG~1216+069, PG~1411+442, PG~1309+355 and PG~1425+267, sources which have been labelled ``X-ray weak'' by Laor et al. (1997), shown to contain a warm absorber (e.g. Ashton et al. 2004; Reeves \& Turner 2000) or for which the broken power law provides an inadequate fit to the spectrum.

\begin{table*}
\caption{Spearman rank order correlation coefficients ($\rho \ge 0.5$ to 99\% confidence) calculated for each of the scatter plots in Figures~\ref{fig:optical-pi} and \ref{fig:optical-xray}. Optical/UV luminosities are from Laor et al. (1997) and converted into the WMAP equivalent. For each pair of parameters, the first value of $\rho$ is for the plots from Figures~\ref{fig:optical-pi} and \ref{fig:optical-xray}. The second value is obtained through further omission of the five discrepant objects; these quasars may be discrepant on account of being an ``X-ray weak quasar'' and containing a warm absorber (PG~1001+054, PG~1309+355, PG~1411+442 and PG~1425+267.) or on account of being less-well fit by a broken power-law than the majority of other sources (PG~1216+069). We note that a number of these correlation coefficients can be increased still further through omission of {\em only} PG~1001+054 and PG~1411+442, but do not include these values for consistency with Fig.~\ref{gammagamma}.}  
\begin{tabular}{lcccc}
\hline
\hline
                      &Luminosity (3000~\AA)&Luminosity ({\sc H}$\beta$)&Luminosity ([{\sc oiii}])&FWHM ({\sc H}$\beta$)\\
\hline
Low Energy $\Gamma$   &--                   &--                         &--/$-0.66$             &$-0.53$/$-0.86$       \\
High Energy $\Gamma$  &--                   &--                         &$-0.52$/$-0.60$        &$-0.52$/$-0.79$         \\
Break Energy          &--                   &--                         &--                     &--                   \\
\hline
Luminosity (0.3 keV)  &0.59/0.69            &0.52/0.48                  &--                     &--                   \\
Luminosity (2.0 keV)  &0.77/0.77            &0.82/0.73                  &0.63/0.65              &0.55/0.45            \\
Luminosity (10.0 keV) &0.67/0.58            &0.78/0.66                  &0.74/0.78              &0.78/0.73            \\
\hline
\hline
\multicolumn{5}{l}{Low Energy $\Gamma$ vs. High Energy $\Gamma$: $\rho \sim0.97$ to 99\% confidence (see text and Fig.~\ref{fig:gammagamma})}\\
\hline
\end{tabular}
\label{tab:correlation}
\end{table*}

\begin{figure*}
\begin{center}
\leavevmode
\epsfig{file=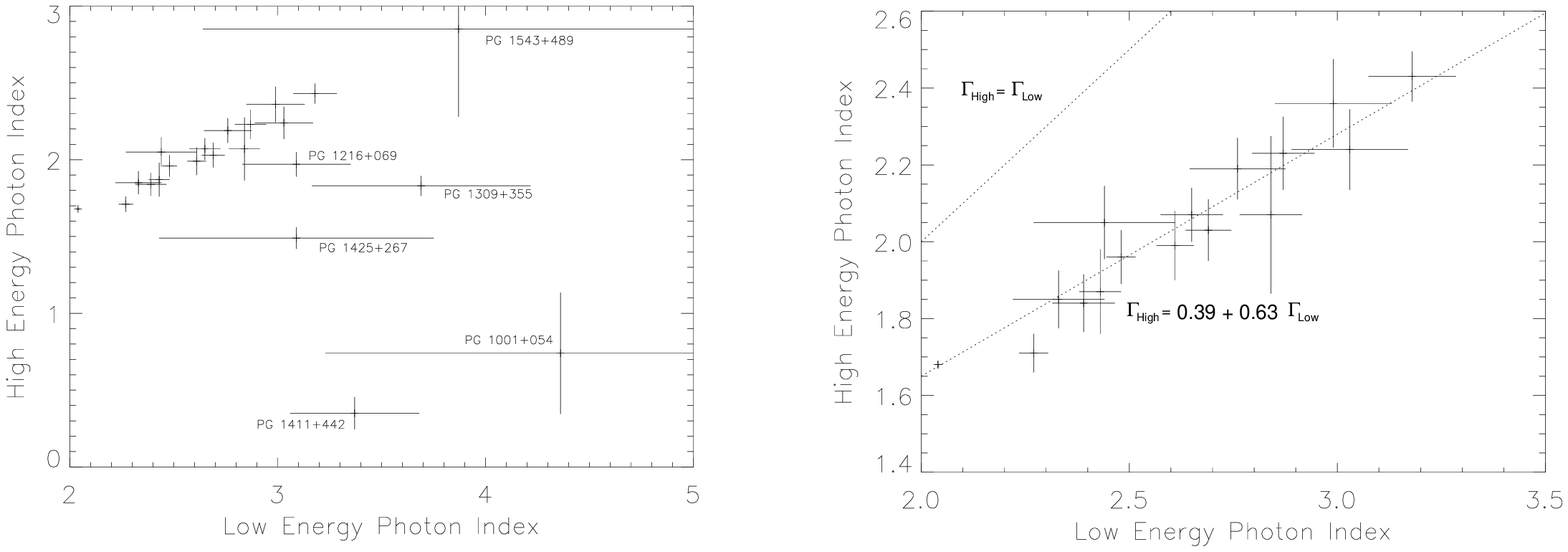, width=18cm, angle=0}
\caption{LHS: Scatter plot showing the high energy photon indices plotted against the low energy photon indices. Clearly, with the exception of the six discrepant objects, there is a strong correlation ($\rho \sim0.97$ to $99\%$ confidence). This suggests that the regions responsible for the emission are either one and the same or in direct ``contact'' with each other. The discrepant objects are mainly those which contain warm absorbers (Ashton et al. 2004) and/or are defined as ``X-ray weak'' (Laor et al. 1997). PG~1216+069 is poorly-fit by the broken power-law but it is less clear why it should be discrepant. RHS: Expanded view of the highly-correlated region of the plot. The dotted lines correspond to $\Gamma_{High}=\Gamma_{Low}$ and $\Gamma_{High}=0.39+0.63\Gamma_{Low}$, where the errors on the intercept and gradient are 0.08 and 0.04 respectively.} 
\label{fig:gammagamma}
\end{center}
\end{figure*}

Allowing for the few discrepant sources which require a more complicated model, it is clear that there are a number of strong correlations present. In Figure~\ref{fig:optical-pi} we show that both the hard and soft photon indices are correlated with the [{\sc oiii}] luminosity and the {\sc H}$\beta$ FWHM . The break energies do not show any form of correlated behaviour with other quantities. Similarly the 3000~\AA~and {\sc H}$\beta$ luminosities do not seem to show any dependency on the X-ray spectral parameters. Figure~\ref{fig:optical-xray} shows that the 2.0 and 10.0 keV luminosities are clearly correlated with all four optical parameters; the 0.3 keV luminosities are not correlated with any of the optical parameters. 

These results confirm the results of Laor et al. (1997), Reeves \& Turner (2000) and Brandt, Mathur \& Elvis (1997) who find similar correlations, most notably the result that the X-ray spectrum steepens as the FWHM of the {\sc H}$\beta$ line decreases. Reeves \& Turner (2000) suggest that, since objects with narrower {\sc H}$\beta$ lines are thought to accrete at a higher fraction of the Eddington limit, there is greater Compton cooling of the hard X-ray-emitting corona and thus a steepening of the spectrum. They also suggest that a soft excess component associated with the accretion disc should therefore be present in the objects with narrower lines. However, we find that the soft excess is actually present below 2 keV in {\em all} sources in the Laor et al. sample which do not contain some sort of warm absorber. Instead we compare the photon indices of the low and high energy power-laws used in our broken power-law fits and find a strong correlation, confirming the result of Porquet et al. (2004); $\rho=0.97$ (to within $>99\%$ confidence) is obtained when we omit the (consistently discrepant) sources PG~1001+054, PG~1411+442, PG~1216+069, PG~1309+355 and PG~1425+267.  We note that the correlation is stronger for our sample than for that of Porquet et al. (2004), who obtain $\rho=0.85$; this may be a reflection either on the slightly different sample or on our use of the broken power-law, instead of fixed energy ranges in the observed frame. We discuss the implications of this correlation further in Section 6.

\section{Spectral Energy Distributions}

The most comprehensive compilations of AGN/quasar spectral energy distributions (SEDs) presented to date are those of Elvis et al. (1994) and Kuraszkiewicz et al. (2003). These authors included 2--10 keV X-ray data obtained by the {\sl HEAO} missions, archive data obtained from the literature and Kuraszkiewicz et al. (2003) also presented far-infrared {\sl ISO} data. In particular, the Kuraszkiewicz et al. (2003) sample is hard X-ray selected which is unbiased by effects due to obscuration and reprocessing.

The Laor et al. (1997) sample was optically-selected with the intent of defining a representative sample of X-ray properties; however we note that this has the effect of biasing the sample against obscured sources, as discussed by Kuraszkiewicz et al. (2003). Nonetheless, the quality of the {\sl XMM-Newton} data that we present here is sufficiently superior to the {\sl HEAO} data that compiling SEDs is a valuable exercise, even if the multiwavelength results cannot be considered representative of all quasars.

In Fig.~\ref{fig:nufnu} we plot the optical--X-ray spectra (OM photometry plus EPIC spectra) for each quasar in the Laor et al. sample. They have been plotted in $\nu F_{\nu}$ space in Fig.~\ref{fig:nufnu} and have been corrected for Galactic absorption. Clearly the plots show a wide range of spectral shapes and we discuss this in detail below.

\begin{figure*}
\begin{center}
\leavevmode
\hspace*{-1cm}\epsfig{file=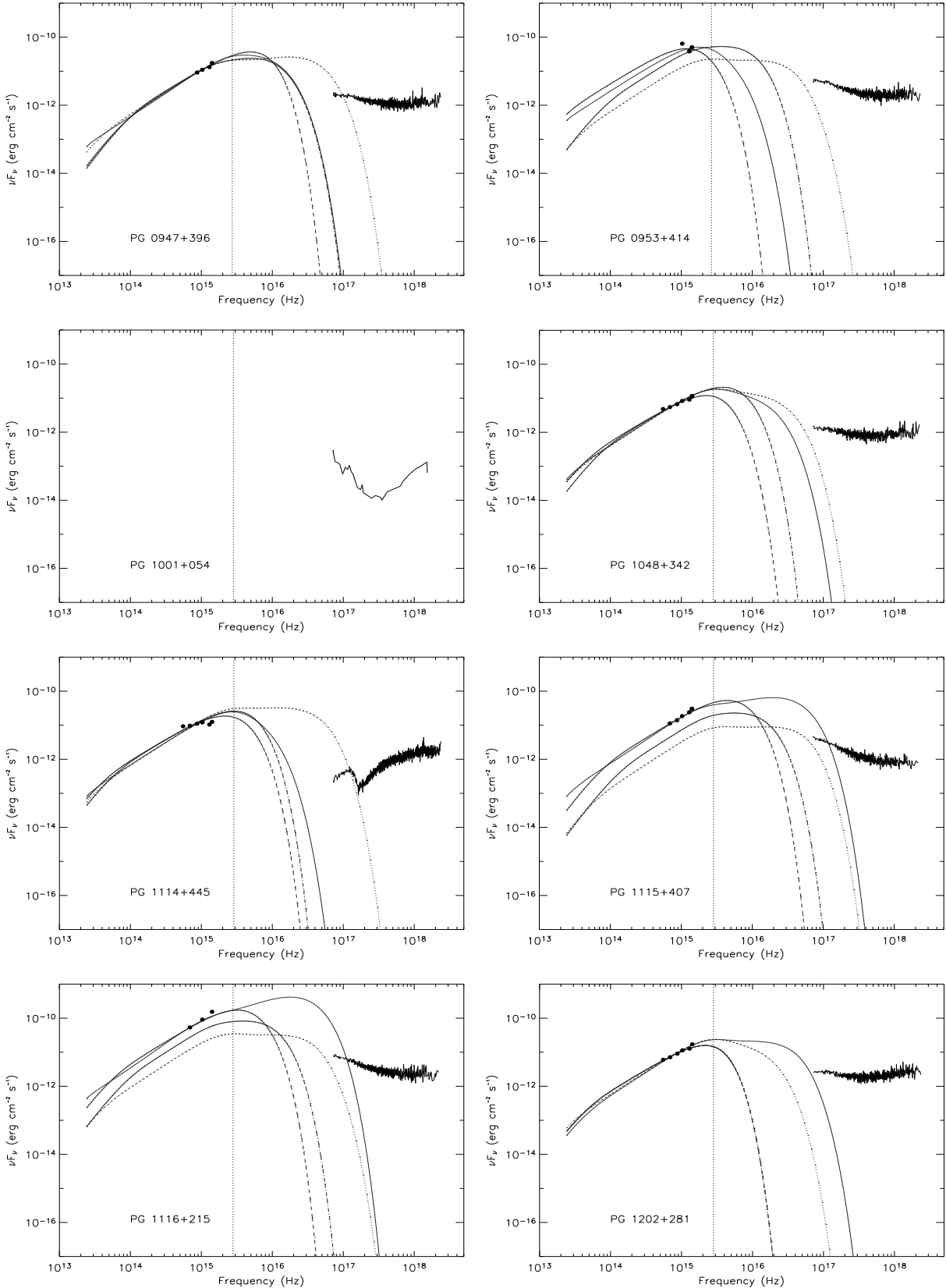, width=16cm}
\caption{Broad-band optical--X-ray spectra (in the observed frame) for each quasar in the Laor et al. sample. We include the {\sl XMM-Newton} X-ray spectra (lines) and OM photometry (filled circles) studied in this paper. All data have been corrected for Galactic absorption, using the extinction curves of Cardelli et al. (1989) where appropriate. The vertical dotted line represents the Lyman limit for each quasar. The best-fit Kerr (solid line) and Schwarzschild (dashed line) Comptonized accretion disc models have been over-plotted, as have the published-mass Kerr (dotted line) and Schwarzschild (dot-dashed line) models.}
\label{fig:nufnu}
\end{center}
\end{figure*}

\begin{figure*}
\begin{center}
\leavevmode
\setcounter{figure}{4}
\hspace*{-1cm}\epsfig{file=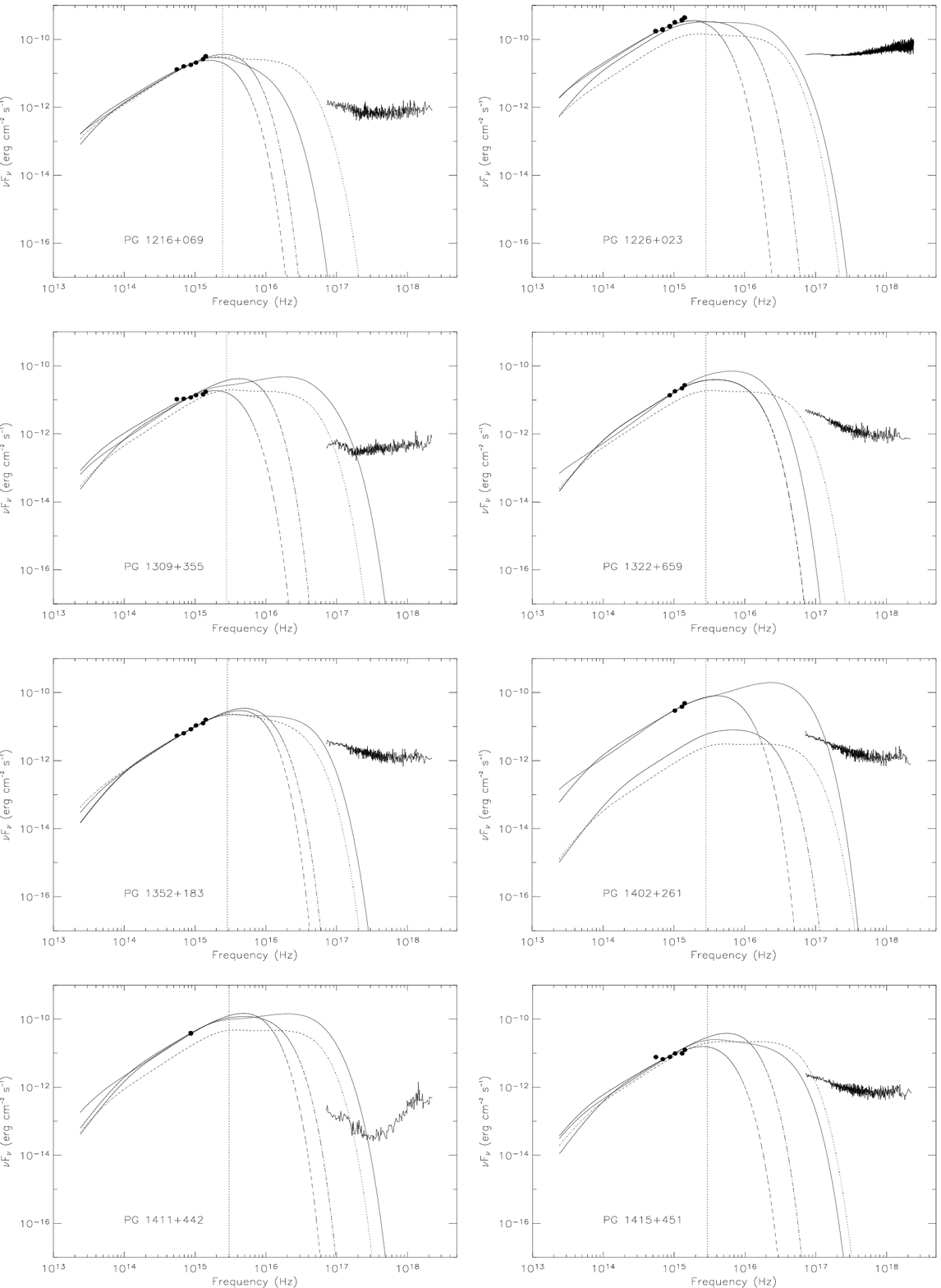, width=16cm}
\caption{(Cont.)}
\label{fig:nufnu}
\end{center}
\end{figure*}

\begin{figure*}
\begin{center}
\leavevmode
\setcounter{figure}{4}
\hspace*{-1cm}\epsfig{file=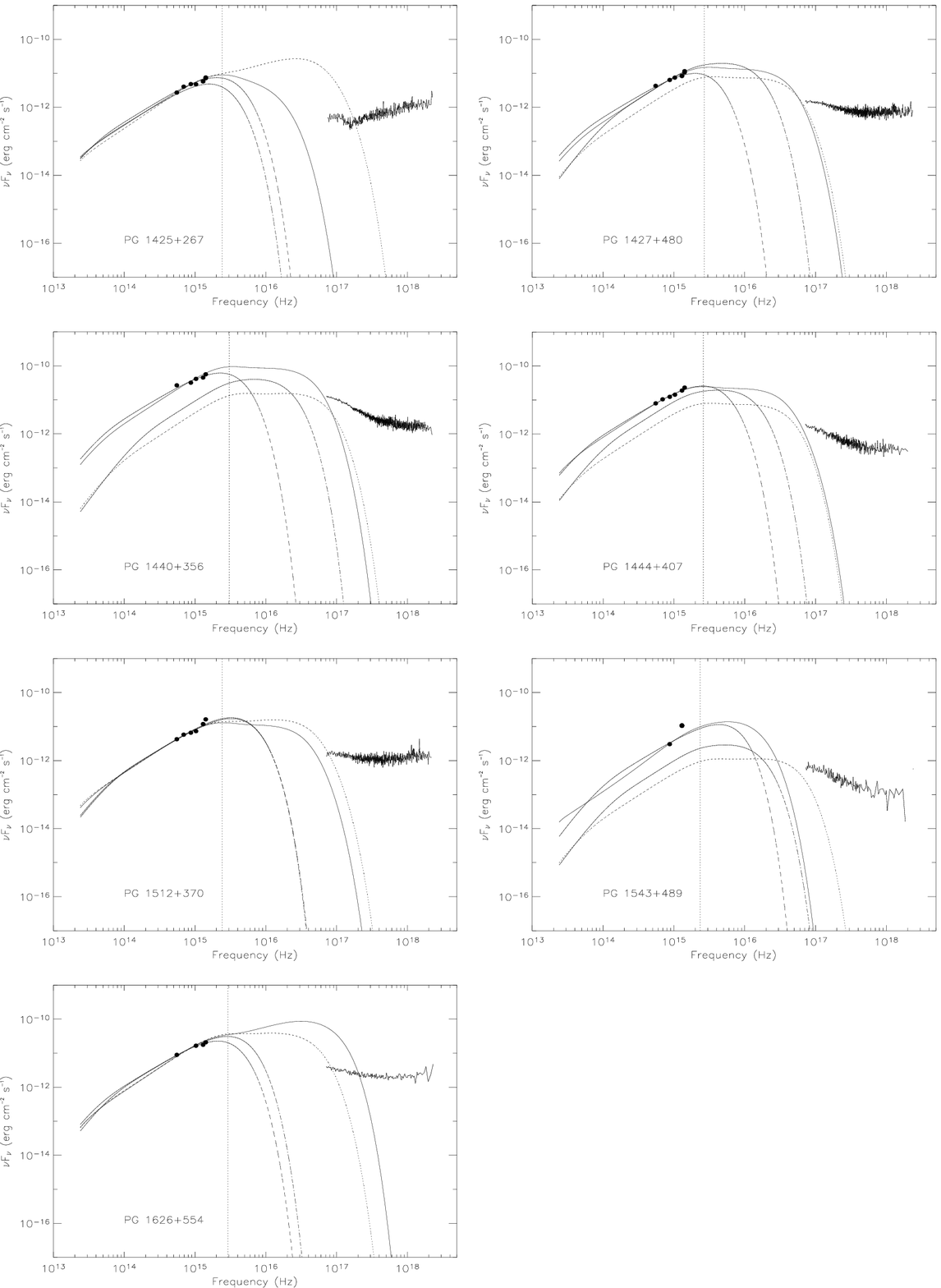, width=16cm}
\caption{(Cont.)}
\label{fig:nufnu}
\end{center}
\end{figure*}

The SEDs are dominated, as expected (e.g. Elvis et al. 1994), by the BBB in the optical--UV region. The turnover of the BBB is not observed for any of the Laor et al. sample. However, as already noted, the majority of the X-ray spectra show a soft excess. Analysis of {\sl ROSAT} data by Puchnarewicz et al. (1996 and references therein) revealed evidence for correlations between the optical and X-ray spectral slopes and the optical:X-ray flux ratio. Such correlations led to the conclusion that there is a strong behavioural link between the BBB and the soft X-ray excess. The additional correlations reported in Section 4 and references therein support this conclusion. The correlation between the soft and hard X-ray power-law slopes suggests further still that there must be a behavioural link between the BBB and the X-ray spectrum as a whole, as opposed to the soft excess in isolation.

We have fit our XMM-OM photometry with Comptonized accretion disc curves generated by a modified Czerny \& Elvis (1987) model (Siemiginowska et al. 1995; Starling et al. 2004). The accretion disc spectra are calculated assuming that the disc is geometrically thin, emitting locally as a modified blackbody (including electron scattering and Comptonization in the atmosphere of the disc) in a Kerr geometry. They do not include the effects of irradiation but are appropriate for simple modelling of the soft excess of X-ray spectra. The best-fit models are plotted on top of the data in Fig.~\ref{fig:nufnu} (solid lines) and appear to be a good or reasonable representation of the OM data for a number of objects. The values of $\chi^2_{\nu}$ are listed with the best-fit masses in Table~\ref{tab:sed-fit}; while some values of $\chi^2_{\nu}$ are relatively high we note that for this preliminary study of the SEDs we have used only a conservative estimate (optical: 1\%, UV: 5\%) for the flux calibration systematic errors and have relatively few points available for constraining the fits. The results also include estimates of the mass accretion rate and inclination angle; the full range of parameter-space is sampled in each case and the results also listed in Table~\ref{tab:sed-fit}. We inclulde approximate error ranges (to within 99\% confidence) for the masses but were unable to constrain the mass accreion rate, luminosity or inclintaion angle. In order to determine the errors we have obtained the volume in parameter-space that is contained within $\Delta\chi^2$ of the minimum $\chi^2$ for each fit. These error ranges should, however, be used with caution since they do not reflect exactly those points of parameter-space which were included within the $\Delta\chi^2+\chi^2_{\min}$ volume.

It is clear from Fig.~\ref{fig:nufnu} that, despite reasonable fits to the OM data, the Kerr models often predict an excess of flux compared with the observed X-ray data (6 out of 22 cases). While we might expect this for those sources known to contain warm absorbers, this is not a satisfactory explanation for the remainder (e.g. PG~1115+407). If instead we use the published masses from the literature (obtained by line width techniques; Shields et al. 2003; Vestergaard 2002; Baskin \& Laor 2005), the fits are much worse ($>>3\sigma$ in most cases) and the predicted soft X-ray flux exceeds that observed in 7 systems. Most of the masses we derive, by fitting the OM data, are higher than published estimates in the literature; accretion rates in excess of the Eddington limit (and outside the model limits) and/or the effects of e.g. irradiation may be required in order to provide acceptable fits to these published mass values. In 11 objects (e.g. PG~0953+414), the models that use the published masses (dotted lines in Fig.~\ref{fig:nufnu}) underestimate the optical/UV flux compared with the OM points. This is somewhat worrying since it shows that either the models are inadequate and e.g. irradiation needs to be included, or it is incorrect to assume that there is no significant contribution to the optical flux from e.g. host galaxy contamination and/or emission lines. Alternatively it may suggest that the source luminosity has varied since the mass estimates were published, as would be consistent with the discrepancies between the OM and IUE data in some objects. Finally, we do not confirm the correlation between mass and 2--10 keV luminosity ($\rho\sim0.5$ to within 99.7\% confidence) as found by Piconcelli et al. (2004), using either our derived masses or the published masses of the Laor PG quasars.

The most striking feature of these plots is that there appears to be no relationship between the fits and the X-ray spectra. There is no source for which the soft excess has a sufficiently steep spectrum that it can be reproduced by the disc model, in contrast with a source like Mkn~841 for which a disc spectrum can be fit to the combined soft excess and UV spectrum (Arnaud et al. 1985). Indeed, we find that the slope and flux of the soft excess is not predicted by the disc model to any significant degree, which provides an interesting challenge for interpreting the optical/UV/X-ray correlations discussed above.

As an alternative, we have also performed fits to the spectra using Schwarzschild models, both with and without modifications to correct for the effects of opacity. The values of $\chi^2_{\nu}$ marked with ``m'' in Table~\ref{tab:sed-fit} indicate which of the best-fit Schwarzschild models had been modified to include the effects of opacity. These fits are shown in Fig.~\ref{fig:nufnu} (dashed lines). Schwarzschild models do not predict significant flux in the X-ray region, and thus the problem of overestimating the X-ray flux is eliminated. Table~\ref{tab:sed-fit} shows that the fits to the optical/UV data are improved for the majority of the Laor objects, compared with the Kerr fits ($\Delta\chi^2>1.0\sigma$ for 17 objects and $\Delta\chi^2>2.6\sigma$ for 11 objects). There is also a marginal increase in agreement between the masses derived and the published values (dot-dashed lines in Fig.~\ref{fig:nufnu}). 

Similar modeling of sub-sets of the Laor et al. sample was performed by other authors; we find that our resultant masses are comparable with, although better-constrained than, those of Laor (1990) but are consistently lower than those of Sun \& Malkan (1989). We also note that Laor (1990) found that Kerr models were more successful at describing the data. Whilst there is evidence for a rotating black hole in MCG--6-30-15 (Wilms et al. 2001; Vaughan \& Fabian 2004) and broad iron lines are seen in several more AGN for which a relativistic disc line is a viable explanation (e.g. Mkn 766, Mason et al. 2003; Q0056-363, Porquet \& Reeves 2003; Mkn 205, Reeves et al. 2001), our preliminary study of the broad-band SEDs of Laor PG quasars might suggest that a Schwarzschild geometry is more appropriate. In any case the soft X-ray excess appears to be independent of the disc model.

\begin{table*}
\caption{Results from fitting the Kerr and Schwarzschild SED models to the OM photometry. The first two columns list the sources and the number of degrees of freedom for the fits. The next four columns (first line of data) show the best-fit mass and $\chi^2_{\nu}$ of the Kerr and Schwarzschild models respectively. The seventh column lists the published mass-estimates for each source (from $^a$Shields et al. 2003, $^b$Vestergaard 2002; $^c$Baskin \& Laor 2005) and the final two columns show the minimum $\chi^2_{\nu}$ obtainable for these published masses. Derived masses for which 99\% confidence intervals include the published mass values are marked by an asterisk. The values of $\chi^2_{\nu}$ marked with ``m'' indicate which of the best-fit Schwarzschild models had been modified to include the effects of opacity. We find that the values of $\chi^2_{\nu}$ are relatively high in some cases, but the majority appear to provide reasonably successful representations of the OM data when plotted in Fig.~\ref{fig:nufnu}. In approximately 50\% of the sources, (typically those for which $\chi^2_{\nu} \le 13$) the fits have a probability $\ge0.05$ that the model is an acceptable representaion of the data. The second line of data for each source gives the fractional mass accretion rate (Kerr models), fractional luminosity (Schwarzschild models) and the cosine of the inclination angle -- the models sampled the full parameter space of these parameters and we include their values for completeness. N.B. In order to determine errors (to within 99\% confidence) on the resultant masses, we have obtained the volume in parameter-space that is contained within $\Delta\chi^2$ of the minimum $\chi^2$ for each fit. This does not enable us to constrain the values of mass accretion rate, luminosity or inclination angle but gives us approximate error ranges for the masses. These error ranges should, however, be used with caution since they do not reflect exactly those points which were included within the $\Delta\chi^2+\chi^2_{min}$ volume of parameter-space.}
\hspace*{-1cm}\begin{tabular}{lcccccccc}
\hline
\hline
&&\multicolumn{2}{l}{Best-Fit Kerr}&\multicolumn{2}{l}{Best-Fit Schwarzschild}&\multicolumn{1}{l}{Published-Mass} &  &\\
Quasar&D.O.F.      &BH Mass ($\log M_{\odot}$)           &$\chi^2_{\nu {\rm ,\,Kerr}}$&BH Mass ($\log M_{\odot}$)           &$\chi^2_{\nu {\rm ,\,Schwarz.}}$&BH Mass ($\log M_{\odot}$)           &$\chi^2_{\nu {\rm ,\,Kerr}}$&$\chi^2_{\nu {\rm ,\,Schwarz.}}$\\
&	       &$\dot{M}/\dot{M}_{\rm Edd}\,\,\,\,\cos i$ &        &$L/L_{\rm Edd}\,\,\,\,\cos i$             &            &           & $\dot{M}/\dot{M}_{\rm Edd}\,\,\,\,\cos i$             &$L/L_{\rm Edd}\,\,\,\,\cos i$	\\
\hline
PG 0947+396&4  &9.1$^{+0.1}_{-0.7}$*     & 1.6	 & 8.1$^{+0.7}_{-0.0}$*       	&  1.6   &   8.54$^a$         &       2.9        &    2.1(m)  \\ 	    
&	       &0.10 0.50  	        &        & 1.00 1.0 	                &        &                    &0.80 0.75         & 1.0  0.25           \\ 
PG 0953+414&4  &9.4$^{+0.2}_{-0.5}$     &20.6	 & 9.8$^{+0.0}_{-0.5}$     	& 15.6(m)&   8.51$^a$         &     174.3        &   33.8(m)  \\ 	 
&	       &0.10 1.00  	        &        & 0.25 0.1 	                &        &                    &0.80 1.00         &1.0  1.00            \\
PG 1001+054&-- &--       	        &  --    & --      	                &        &   7.65$^a$         &         --       &      --    \\ 	 
&	       &--       	        &  --    &         	                &        &                    &                  &            \\ 
PG 1048+342&6  &8.4$^{+0.3}_{-0.3}$*    & 8.0	 & 8.5$^{+0.0}_{-0.1}$     	& 3.3(m) &   8.25$^a$         &       8.3        &    7.0     \\ 
&	       &0.20 1.00  	        &        & 0.75 0.1 	                &        &                    &0.30 1.00         &0.4  0.50            \\
PG 1114+445&6  &8.6$^{+0.0}_{-0.0}$     &74.5	 & 8.7$^{+0.0}_{-0.0}$     	&  37.0  &   8.41$\pm0.09^b$  &      85.5        &   57.3     \\ 	 
&	       &0.10 1.00  	        &        & 0.50 0.1 	                &        &                    &0.50 0.75         &0.2  0.75            \\
PG 1115+407&5  &8.8$^{+0.2}_{-0.5}$     & 6.8	 & 8.5$^{+0.1}_{-0.5}$     	&   5.9  &   7.505$^c$        &    1207.8        &  627.4(m)  \\ 	 
&	       &0.30 0.50  	        &        & 0.25 0.9 	                &        &                    &0.80 1.00         &1.0  1.00            \\
PG 1116+215&2  &9.7$^{+0.2}_{-0.0}$     &12.2	 & 9.1$^{+0.1}_{-0.4}$     	&  14.6	 &   8.50$\pm0.10^b$  &    2196.7        &  881.0(m)  \\ 	 
&	       &0.30 0.25  	        &        & 0.25 1.0 	                &        &                    &0.80 1.00         &1.0  1.00            \\
PG 1202+281&5  &8.2$^{+0.7}_{-0.0}$*    & 4.0	 & 8.5$^{+0.1}_{-0.4}$*      	&  2.5(m)&   8.12$\pm0.09^b$  &       6.1        &    2.5(m)  \\ 	 
&	       &0.70 1.00  	        &        & 1.00 0.1 	                &        &                    &0.20 1.00         & 0.1  1.00           \\
PG 1216+069&6  &9.3$^{+0.0}_{-0.1}$*     &13.0	 & 9.6$^{+0.0}_{-0.4}$*     	&  10.4  &   9.17$\pm0.15^b$  &      17.0        &   13.4     \\ 	 
&	       &0.20 1.00  	        &        & 0.25 0.2 	                &        &                    &0.70 1.00         & 0.5  0.75           \\
PG 1226+023&28 &9.3$^{+0.0}_{-0.0}$     &20.8	 & 9.7$^{+0.0}_{-0.0}$     	&  18.3  &   8.88$^a$         &    2164.2        &  253.8(m)  \\ 	 
&	       &0.70 1.00  	        &        & 0.25 0.4 	                &        &                    &0.80 1.00         & 1.0  1.00           \\
PG 1309+355&5  &9.0$^{+0.0}_{-0.6}$     &68.4	 & 8.8$^{+0.0}_{-0.1}$     	&  32.6  &   8.21$\pm0.13^b$  &     363.2        &   67.4     \\ 	 
&	       &0.80 0.25  	        &        & 0.75 0.1 	                &        &                    &0.80 1.00         &0.7  1.00            \\
PG 1322+659&3  &9.3$^{+0.0}_{-0.8}$     & 1.0	 & 8.1$^{+0.7}_{-0.0}$      	&  1.8(m)&   8.09$^a$         &     358.3        &    1.8(m)  \\ 	 
&	       &0.10 0.25  	        &        & 1.00 0.8 	                &        &                    &0.80 1.00         & 0.8  1.00           \\
PG 1352+183&5  &8.1$^{+0.7}_{-0.0}$*    & 3.1	 & 8.0$^{+0.5}_{-0.2}$*      	&   2.1  &   8.27$\pm0.09^b$  &       3.3        &    3.0     \\ 	 
&	       &0.70 1.00  	        &        & 1.00 0.5 	                &        &                    &0.30 1.00         &0.9  0.25            \\
PG 1402+261&2  &9.3$^{+0.0}_{-0.8}$     & 2.7	 & 8.7$^{+0.3}_{-0.5}$     	&   2.7  &   7.845$^c$        &     490.0        &  455.1(m)  \\ 	 
&	       &0.30 0.25  	        &        & 0.25 1.0 	                &        &                    &0.80 1.00         &1.0  1.00            \\
PG 1411+442&6  &8.6$^{+0.4}_{-0.2}$     & 1.4	 & 8.1$^{+0.3}_{-0.1}$     	&   0.6  &   7.87$^a$         &    3139.2        &   22.7(m)  \\ 	 
&	       &0.60 0.50  	        &        & 0.75 0.8 	                &        &                    &0.80 1.00         &1.0  1.00            \\
PG 1415+451&5  &8.0$^{+0.5}_{-0.0}$     &55.8	 & 8.4$^{+0.0}_{-0.0}$     	&  35.8  &   7.81$\pm0.09^b$  &      99.8        &   49.1     \\ 	 
&	       &0.40 1.00  	        &        & 0.50 0.1 	                &        &                    &0.80 1.00         &0.7  0.75            \\
PG 1425+267&7  &8.9$^{+0.4}_{-0.3}$*     & 6.5	 & 9.2$^{+0.2}_{-0.0}$*     	&   3.8  &   9.317$^c$        &       7.8        &     43.5(m)\\ 	 
&	       &0.20 1.00  	        &        & 0.25 0.2 	                &        &                    &0.50 0.25         &0.1  0.25            \\
PG 1427+480&6  &8.3$^{+0.7}_{-0.0}$     & 4.8	 & 8.7$^{+0.0}_{-0.5}$     	&  2.6(m)&   7.99$^a$         &     223.6        &    7.4(m)  \\ 	 
&	       &0.70 1.00  	        &        & 0.75 0.1 	                &        &                    &0.80 1.00         & 1.0  1.00           \\
PG 1440+356&4  &8.1$^{+0.1}_{-0.0}$     &95.6	 & 8.5$^{+0.0}_{-0.1}$     	&  51.1  &   7.30$\pm0.10^b$  &    2442.3        & 1854.2(m)  \\ 	 
&	       &0.70 1.00  	        &        & 0.75 0.1 	                &        &                    &0.80 1.00         &1.0  1.00            \\
PG 1444+407&5  &8.7$^{+0.4}_{-0.0}$     & 5.1	 & 9.1$^{+0.0}_{-0.6}$     	&   4.1  &   8.23$\pm0.10^b$  &     995.2        &  328.2(m)  \\ 	 
&	       &0.70 1.00  	        &        & 0.25 0.4 	                &        &                    &0.80 1.00         &1.0  1.00            \\
PG 1512+370&5  &8.8$^{+0.7}_{-0.0}$*    &12.6	 & 8.8$^{+0.3}_{-0.3}$*      	&  10.9  &   8.95$\pm0.10^b$  &      14.2        &   12.6     \\ 	 
&              &0.60 1.00  	        &        & 0.50 0.7 	                &        &                    &0.50 0.75         &0.8  0.25            \\
PG 1543+489&5  &9.5$^{+0.0}_{-0.0}$     &49.5	 & 8.3$^{+0.5}_{-0.0}$     	&  58.4  &   7.844$^c$        &      573.1       &   420.9 (m)\\ 	 
&	       &0.10 0.25  	        &        & 1.00 1.0 	                &        &                    &0.80 1.00         & 1.0  1.00           \\
PG 1626+554&3  &8.8$^{+0.0}_{-0.6}$*    &12.5	 & 8.7$^{+0.2}_{-0.4}$*      	&  1.2(m)&   8.37$^a$         &      15.4        &    4.3     \\ 	 
&	       &0.80 0.25  	        &        & 0.50 0.1 	                &        &                    & 0.50 0.75 	 &0.2  0.75            \\
\hline
\end{tabular}
\label{tab:sed-fit}
\end{table*}

\section{Discussion}

\begin{figure}
\begin{center}
\leavevmode
\hspace*{-1cm}\epsfig{file=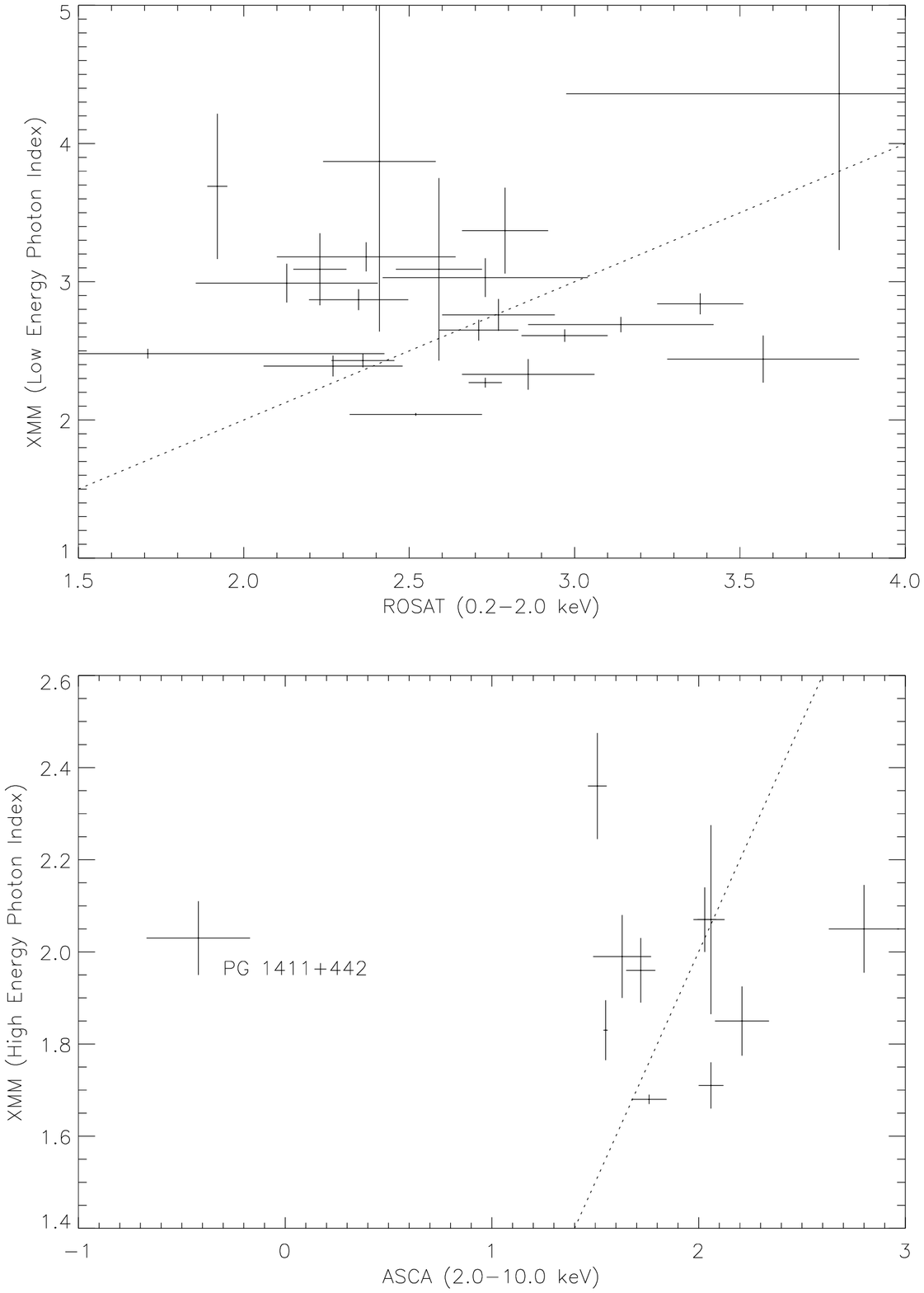, width=8cm}
\caption{Comparison of the {\sl XMM} photon indices with those of {\sl ROSAT} (Laor et al. 1997; the energy indices quoted have been converted to photon indices) and {\sl ASCA} (George et al. 2000; Reeves \& Turner 2000). The dotted lines represent the point at which each pair of telescopes would obtain equivalent results.}
\label{satellites}
\end{center}
\end{figure}

The BQS is a well-studied sample and so there are various results in the literature with which to compare the {\sl XMM-Newton} spectra. In particular, Laor et al. (1997) obtained values of the photon index for the 1.2--3.0 keV range which are comparable with the values we obtain for the softer of the two power laws, i.e. $\Gamma =$ 2.2--3.7 below the break energy. We note that our derived break energies are in the range 0.9--2.6 keV, whereas in effect the Laor et al. study was equivalent to assuming a break energy $>$2 keV, and so this a reasonable but not exact comparison\footnote{For simplicity and comparison with the literature, we discuss the spectra in terms of their power-law parameters but note that the successful fitting of e.g. blackbody and Comptonization models to the soft X-rays by other authors may indicate a curved spectrum instead of a power-law.}. Comparison of the {\sl XMM} and {\sl ROSAT} photon indices is shown in the top panel of Fig.~\ref{satellites}.

The BQS sample has also been observed previously at harder energies using the {\sl ASCA} satellite. George et al. (2000) found a range $\Gamma =$ 1.5--3.0 which is consistent with the {\sl XMM-Newton} values for the Laor et al. sample. PG~1411+442 was the only source in the George et al. (2000) sample that could not be modelled successfully with a single power law in this range and, similarly, we have found that it warrants more detailed fitting than performed in this paper. {\sl ASCA} data from a larger sample of 62 quasars was studied by Reeves \& Turner (2000) and similar values obtained for $\Gamma$ in the 2--10 keV range.  Comparison of the {\sl XMM} and {\sl ASCA} photon indices is shown in the bottom panel of Fig.~\ref{satellites}.

Table~\ref{tab:summary} provides a summary of X-ray properties of individual sources. In only four of the quasars studied in this paper -- PG~1216+069, PG~1309+355, PG~1425+267 and PG~1440+356 -- could an additional neutral absorption component be detected. Warm absorbers have been discovered in the {\sl XMM-Newton} spectra of PG~1114+445 and PG~1309+355 (Ashton et al. 2004) and possibly PG~1425+267 and PG~1411+442 (Ashton, Brocksopp et al. in prep.). These findings are consistent with Laor et al. (1997), who identified a distinct group of ``X-ray-weak quasars'' consisting of PG~1001+054, PG~1411+442 and PG~1425+267 on account of their absorption properties. We note that these same X-ray-weak objects are among the sources which show signs of UV absorption in their spectra (Laor \& Brandt 2002) Reeves \& Turner (2000) found absorption in just over half of their sources, mainly in the more distant objects but also in some of the low-redshift radio-loud quasars. In particular they discuss possible warm absorber components in PG~1226+023 and PG~1425+267, again consistent with our results. This fraction of Laor sources with a warm absorber is notably less than the 50\% of a sample of Seyfert 1 galaxies (Reynolds 1997), as discussed by Laor et al. (1997) and Ashton et al. (2004).

We find little evidence for iron lines in the {\sl XMM-Newton} spectra although the upper limits to the equivalent width allow for appreciable line emission in some other sources. Detections for which the equivalent width could be constrained (out to 99\% confidence) were obtained for just twelve sources but to low significance ($<2\sigma$). Our upper limits are consistent with the {\sl ASCA} results since Reeves \& Turner (2000) report iron line detection in PG~1114+445, PG~1116+215, PG~1226+023 and PG~1425+267. We further note that Page et al. (2004a) detect weak iron emission in PG~1226+023 when they combine a series of nine observations together. Similarly our results are consistent with those of Porquet et al. (2004).

\begin{table*}
\caption{Summary of the X-ray properties of each individual PG quasar.}
\hspace*{-1cm}\begin{tabular}{ll}
\hline
\hline
PG 0947+396&Good fit to broken power-law\\
PG 0953+414&Good fit to broken power-law\\
PG 1001+054&Apparent good fit to broken power-law (due to low count rate?) but discrepant parameters, ``X-ray weak quasar''\\
PG 1048+342&Good fit to broken power-law\\
PG 1114+445&Poor fit to broken power-law, warm absorber present\\
PG 1115+407&Good fit to broken power-law, NLS1\\
PG 1116+215&Good fit to broken power-law\\
PG 1202+281&Good fit to broken power-law\\
PG 1216+069&Neutral absorption component required, although poor fit to broken power-law\\
PG 1226+023&$\equiv$ 3C~273, radio-loud, highly luminous, neutral absorption component required although poor fit to broken power-law\\
PG 1309+355&Radio-loud, warm absorber present, good fit to broken power-law\\
PG 1322+659&Good fit to broken power-law\\
PG 1352+183&Good fit to broken power-law\\
PG 1402+261&Reasonable fit to broken power-law\\
PG 1411+442&Poor fit to broken power-law, absorption and higher-energy component required, ``X-ray weak quasar''\\
PG 1415+451&Radio-loud, good fit to broken power-law\\
PG 1425+267&Radio-loud, warm absorber probably present, good fit to broken power-law but discrepant soft photon-index, ``X-ray weak quasar''\\
PG 1427+480&Good fit to broken power-law\\
PG 1440+356&Reasonable fit to broken power-law, neutral absorption required, NLS1\\
PG 1444+407&Good fit to broken power-law\\
PG 1512+370&Good fit to broken power-law\\
PG 1543+489&Broken power-law provides good fit but poorly constrained, NLS1\\
PG 1626+554&Good fit to broken power-law\\
\hline
\end{tabular}
\label{tab:summary}
\end{table*}

On fitting a power-law to the 2--5 keV range for each source we find that there is a soft excess in each case where we do not find evidence for a warm absorber. This is in contrast to George et al. (2000) and Reeves \& Turner (2000) who find a soft excess in just a few of their low-redshift quasars, although this is likely to be a reflection on the calibration of {\sl ASCA} at low energies. Porquet et al. (2004) find more evidence for a soft excess component but, as also discussed by Page et al. (2004b), the derived black-body temperatures are thought to be too hot to be direct emission from a thin accretion disc. Instead they consider that UV disc photons are upscattered to X-ray energies via a Comptonizing medium. This is consistent with the correlations we find in Section 4 -- the strong correlation between the soft and hard power-law photon indices would suggest that the 0.3--10 keV emission is all produced by the same Comptonizing medium and not directly by the disc. In this case the optical/X-ray continuum luminosity correlations would also be expected since an increase in accretion rate could lead to an increase in luminosity at all frequencies.

It should therefore be possible to fit a Comptonized disc model to the optical-UV-X-ray region. In Section 5 we attempted to do this using the Czerny \& Elvis (1987) model (both Kerr and Schwarzschild). While our fits would be much better constrained if we were able to sample the peak of the emission (presumably in the EUV range), some of the published mass estimates produce curves which are considerably less luminous than the observed optical/UV data.

Aside from the mass estimates, however, it is clear that the accretion disc models presented here do not account for the soft X-ray excesses of these low redshift quasars. This result, coupled with the correlation between the soft and hard X-ray slopes, leads us to question what sort of physical models {\em are} able to account for the observed spectra. If, as the tight correlation would suggest, the soft and hard X-ray parts of the spectra are produced via the same emission mechanism and in the same emitting region, it is likely that we are looking for a non-thermal mechanism, such as those invoking Comptonization and/or synchrotron, as appears to be the case for the broad-band spectra of X-ray binaries in the low/hard state (e.g. Markoff et al. 2001). Models invoking thermal blackbody and/or bremsstrahlung components are non-physical at high energies and so the correlation would suggest that they are equally inappropriate at low energies. A two-temperature disc-corona model (such as Kawaguchi, Shimura \& Mineshige 2001) requires the low- and high-energy power-law components to be produced in different regions and via different mechanisms, creating difficulty in producing such a tight correlation.

The failure of the accretion disc models to fit the soft X-ray spectrum also poses a possible problem for models of Narrow Line Seyfert 1 (NLS1) galaxies which invoke an accretion disc with large inner radius for the soft X-ray excess (e.g. Pounds, Done, Osborne 1995). With H$\beta$ line widths of 1000--2000 km$\,$s$^{-1}$, PG~1115+407, PG~1440+356 and PG~1543+489 can be classed as NLS1 objects; we find that these sources are more variable and have steeper spectra than most of the other sources in the sample, consistent with other NLS1 objects. However, the models yield considerably larger masses than the 10$^6$--10$^7\,\,M {\odot}$ that would be expected from NLS1 models. Alternative models for these objects might instead invoke a more face-on inclination angle and/or a more distant broad line region (e.g. Boller et al. 1992, Wandel 1997, Puchnarewicz et al. 1992).

We should question whether more complicated multiwavelength models are required. It is known that, like many quasars and active galaxies, PG~1226+023 (3C 273) has jets which emit at radio, optical and X-ray wavelengths (e.g. Courvoisier 1998 and references therein) and it is likely that the other radio-loud PG quasars have similar structures. Some radio-quiet quasars have now had their jets imaged at radio frequencies (e.g. Blundell et al. 2003a) and, although none of the Laor radio-quiet objects have been resolved, they are emitting at significant flux densities of at least a few mili-Janskys. Fitting disc models to any source, for which a jet is making significant contributions to the optical or X-ray luminosity, will be meaningless.

Four of the sources studied here have a significant number of radio points from which we can determine the spectral index; of these two are radio-loud (PG~1425+267, PG~1512+370) and two are radio-quiet (PG~1116+215, PG~1216+069)\footnote{We note however that the concept of ``radio-loud'' and ``radio-quiet'' may have been introduced artificially through obtaining quasar samples according to their optical properties and choosing specific radio observing frequencies (see e.g. Blundell 2003b and references therein) }. Brocksopp et al. (2004) found that for the radio-loud high redshift quasar [HB89] 2000$-$330 the spectral index of the X-rays and radio emission were consistent and suggested that the X-ray emission might share a common origin with the radio, such as synchrotron self-Comptonization of jet photons. We see a similar relationship in just one of the low redshift radio-loud objects studied in this paper, PG~1512+370. In contrast, for the two radio-quiet quasars studied here we find that the spectral indices of the X-rays are steeper than those of the radio. Conversely, the radio-loud quasar, PG~1425+267, has a steeper radio spectrum than X-ray. The radio spectral indices were in the range $-0.6$ to $-0.9$, consistent with optically thin emission. However, we also note that any curvature in the spectra may complicate comparison of the slopes and so a more thorough analysis should include study of the X-ray vs. radio flux correlation. We will address this in a future work.

For these same four radio-loud sources we also use the formulae of Marscher (1983) to estimate the amount of synchrotron self-Compton (SSC) emission expected at X-ray wavelengths, assuming that the radio emission was all emitted in a jet. We find that less than 1\% of the X-ray emission could be from such a source and so an even smaller fraction is to be expected in the radio-quiet sources. Therefore we must look to alternative sources of X-ray emission and more sophisticated jet models in order to account for the observed spectra. For example, Zdziarski (1986) presents a model invoking SSC to explain the IR--X-ray spectra of radio-quiet AGN; similarly the model of Falcke \& Biermann (1995) produces a significant level of SSC at X-ray energies and may be applicable to radio-quiet objects as well as radio-loud. Again, we will be investigating these models in a future work.

\section{Conclusions}
We have analysed {\sl XMM-Newton}/EPIC X-ray spectra for the Laor et al. sample of PG quasars and found that a model invoking a broken power-law and iron line is an acceptable fit in the majority of cases. Those sources for which the model is a poor fit seem to require some sort of warm absorber at low energies and one source appears to need a further higher energy component. These discrepant sources will be studied in more detail in a future paper. Iron lines were detected in a number of sources but with significance $<2\sigma$. We have confirmed correlations between various optical and X-ray line/continuum parameters; in particular we find that the power-law indices of the low and high energy power-laws are well correlated, suggesting that both components have a common origin. We have also analysed {\sl XMM-Newton}/OM photometry and combined this with the X-ray spectra and archive radio--UV photometry in order to compile broadband spectral energy distributions. Fitting these spectra to a Comptonized disc model show that the soft excess appears to be independent of the disc model, adding support to our interpretation of the correlation between the soft and hard X-ray spectra. 

\section{acknowledgments}
We thank Ceri Ashton for useful discussion, Aneta Siemiginowska for the use of her accretion disc models and Gavin Ramsay for proof-reading the manuscript. This is work based on observations obtained with {\sl XMM-Newton}, an ESA science mission with instruments and contributions directly funded by ESA Member States and the USA (NASA). We have made use of the NASA/IPAC Extragalactic Database (NED) which is operated by the Jet Propulsion Laboratory, California Institute of Technology, under contract with the National Aeronautics and Space Administration.

\end{document}